\documentclass[twocolumn]{aastex63}
\usepackage[utf8]{inputenc}
\usepackage{subfigure}
\usepackage{amsmath,amssymb,amsfonts}
\usepackage{color}
\usepackage{lineno}
\usepackage{ulem}

\begin{document}

\title{The fate of EMRI-IMRI pairs in AGN accretion disks: hydrodynamic and three body simulations}

\author{Peng Peng}
\affiliation{Astronomy Department, School of Physics, Peking University, 100871 Beijing, China}

\author{Alessia Franchini}
\affiliation{Department of Astrophysics, University of Zurich, Winterthurerstrasse 190, CH-8057 Z\"urich, Switzerland}

\author{Matteo Bonetti}
\affiliation{Universit\`a degli Studi di Milano-Bicocca, Piazza della Scienza 3, I-20126 Milano, Italy
}
\affiliation{
INFN, Sezione di Milano-Bicocca, Piazza della Scienza 3, I-20126 Milano, Italy
}

\author{Alberto Sesana}
\email{alberto.sesana@unimib.it}
\affiliation{Universit\`a degli Studi di Milano-Bicocca, Piazza della Scienza 3, I-20126 Milano, Italy
}
\affiliation{
INFN, Sezione di Milano-Bicocca, Piazza della Scienza 3, I-20126 Milano, Italy
}

\author{Xian Chen}
\email{xian.chen@pku.edu.cn}
\affiliation{Astronomy Department, School of Physics, Peking University, 100871 Beijing, China}
\affiliation{Kavli Institute for Astronomy and Astrophysics at Peking University, 100871 Beijing, China}

\begin{abstract}
Extreme-mass-ratio inspirals (EMRIs) and intermediate-mass-ratio inspirals (IMRIs) are important  gravitational wave (GW) sources for the Laser Interferometer Space Antenna (LISA). 
It has been recently suggested that EMRIs and IMRIs can both form in the accretion disk of an active galactic nucleus (AGN). Considering the likely encounter between a sBH and an IMBH during the migration in the AGN disk, \citet[][Paper I]{Peng23} showed that a gap-opening IMBH can drive a surrounding sBH to migrate synchronously. In this work, we extend the study in Paper I with a more sophisticated model. We first use 3D hydrodynamical simulations to study the co-evolution of the disk and the migration of a sBH in the vicinity of an IMBH. 
We find that the gaseous torque, together with the tidal torque exerted by the IMBH, can drive synchronized migration until $\sim 10$ Schwarzschild radii from the central supermassive black hole (SMBH). We further use a relativistic three-body code to study the final fate of the sBH in the GW-dominated regime. We find that the sBH can be either captured or kicked out by the IMBH, which will result in either two subsequent IMRIs or an EMRI followed by an IMRI. These events 
will bring rich information about the formation and evolution of sBHs and IMBHs in AGNs.


\end{abstract}

\keywords{Active galactic nuclei (16) --- Intermediate-mass black holes (816) --- Stellar mass black holes (1611) --- Gravitational waves (678)}

\section{Introduction}

\label{sec:intro}

EMRIs and IMRIs are anticipated sources of GW for milli-Hertz (mHz) GW detectors like the Laser Interferometer Space Antenna \citep[LISA,][]{pau18}. An EMRI typically involves a supermassive black hole (SMBH) with $10^6-10^7$ $M_\odot$ and a stellar-mass black hole (sBH) with $10-100 \, M_\odot$, resulting in a mass ratio $q<10^{-4}$. In contrast, IMRIs can form from an IMBH with $10^2-10^4$ $M_\odot$ and a stellar-mass compact object, or from an SMBH and an IMBH, with a mass ratio $q$ in the range $ 10^{-4} \lesssim q \lesssim 10^{-2} $ \citep{2007CQGra..24R.113A}.\footnote{Although the $q=10^{-4}$ threshold is somewhat arbitrary, there are at least two reasons to define EMRIs and IMRIs as separate categories. The first has to do with their astrophysical nature. While EMRIs involve sBHs and SMBHs, which are observationally well established objects, IMRIs involve the observationally elusive IMBHs. The second has to do with their GW modeling. While EMRI can be (at least to some extent) modeled in the point-particle limit, this is not true for the IMRI, which makes it much more complicated to construct accurate wavefroms.} Since the GW radiation timescale is proportional to $q^{-1}$, EMRIs and IMRIs could be detectable in the LISA band for years, accumulating $10^4-10^5$ GW cycles and providing detailed information about the spacetime near a massive black hole \citep{2007CQGra..24R.113A}.

One important place for EMRI formation is the accretion disk of an active galactic nucleus (AGN), which is known to be a possible breeding ground of sBHs \citep[e.g.][]{syer91, artymowicz93, subr99, karas01, levin03, Goodman2004}. Here, sBHs can either form from massive stars in the disk outskirt or be captured from the nuclear star cluster \citep{Goodman2004, syer91}. Once embedded in the disk, sBHs might migrate toward the central SMBH due to hydrodynamical effects such as density wave interactions \citep{artymowicz93, levin03}, thermal torque \citep{Grishin2024}, and head or tail winds in the geometrically thick disk \citep{chakrabarti93, kocsis11, sanchez20}. sBHs reaching close enough to the SMBH will produce EMRIs, with some recent calculations suggesting that these "wet" EMRIs can significantly contribute to the overall cosmic EMRI rates \citep{pan21, pan21b, 2022arXiv220510382D}.

IMRIs can also form in the AGN disks, from disk-embedded IMBHs inspiraling into the central SMBH or IMBHs capturing sBHs. The IMBHs, on the one hand, can be brought into the galactic nucleus through galaxy mergers or inspiraling globular clusters \citep{2014ApJ...796...40M, volonteri03}. Then these IMBHs could be captured by the accretion disk similarly to the capture of sBHs \citep{Ivanov1999}. On the other hand, the IMBHs could also grow from the aforementioned sBHs accumulated in the AGN disk. The growth is due to either the accretion of the surrounding gas \citep{mckernan11,2012MNRAS.425..460M,2021MNRAS.507.3362T,ChenYX2023}, or frequent binary formation \citep[e.g.][]{tagawa20, Li2022_binaryform, Boekholt2023} and merger \citep[e.g.][]{mckernan12, bartos17, stone17, Li2022_binarydisk} with the other sBHs in the disk. Formation and growth of IMBHs is expected to be particularly efficient at specific radii within the AGN disk, known as the ``migration traps'' \citep{bellovary16,secunda19,secunda20, peng21, Grishin2024}, which are typically located at tens to hundreds of Schwarzschild radii ($R_{\rm{S}}$) from the central SMBH. Due to the vanishing gaseous torques around those traps, sBHs will accumulate there, form binaries and finally merge with the assistance of gas and/or a third body. Moreover, the merger remnants will remain in the accretion disk and participate in the next generation of mergers. Through such ``hierarchical mergers'' \citep{2019PhRvL.123r1101Y,2021NatAs...5..749G}, sBHs could gradually grow into the intermediate-mass range. Then these IMBH can form an IMRI by either migrating inward and fall into the central the SMBH, or experiencing another merger with the sBH in the crowded disk. 

Within an AGN disk, sBHs and the IMBHs migrate differentially due to the different driving mechanisms, which could induce encounter between the sBH and the IMBH. For example, an sBH can excite a perturbative gas density wave in the disk, and migrate due to the back reaction, a mechanism called Type-I migration \citep{1980ApJ...241..425G, 1997Icar..126..261W}. In contrast, for the IMBH, the tidal torques on the surrounding gas can be strong enough to open a gap in the disk \citep{ivanov99, gould00}. This gap-opening process is similar to what occurs around gas giants in protoplanetary disks \citep{1979MNRAS.186..799L, 1979MNRAS.188..191L,lin86, artymowicz94}. Such gap-opening IMBH continues to exchange energy and angular momentum with the accretion disk by its tidal force, thus migrating toward the central SMBH \citep{gould00, armitage02, Kanagawa2018}, a process called Type II migration. Type I and Type II migration rates can differ significantly due to the very different mass of the migrating object and different disk structure. So if one has a sBH and the an IMBH, the one with the slower migration rate can be possibly chased up by the faster one, during their migration.


When they encounter, the sBH and the IMBH interacts with each other, which may be analogous to the interaction between a gap-opening gas giant and a sub-gas-opening terrestrial planet or planetesimal in a protoplanetary disk \citep[e.g.][]{1996LPI....27..479H, 2008MNRAS.386.1347P, 2008A&A...482..333P}. In particular, Paper I shows that after the encounter, the migration of the sBH will be largely affected by the IMBH. The IMBH can drive the sBH to migrate synchronously with it, i.e. the sBH is forced to migrate inward with the same migration rate as the IMBH. Such a synchronized migration is driven by the strengthened gaseous torque near the gap \citep{Peng23}, as well as the tidal torque of the IMBH \citep[e.g.][]{Yang2019}. The sBH can migrate synchronously with the outside IMBH until they reach a distance of about $10 \, R_{\rm{S}}$ from the central SMBH, where the sBH and the IMBH can form an EMRI and an IMRI with the central SMBH respectively \citep{Peng23}. Since the EMRI and the IMRI in this scenario are close to each other, this kind of GW source could induce very different phenomena in the LISA detection compared to a single EMRI or IMRI, because of resonances, ejections, chaotic behaviour and other exotic phenomena.

The fate of the sBH is interesting because it affects the type of GW phenomena that might appear in the LISA detector. However, in Paper I, we could not assess this fate; i.e. whether the sBH is kicked out, it is captured by the IMBH, or eventually falls into the SMBH. This was caused by the limited capacity of the N-body code used in Paper I, as well as the semi-analytical 1D model used to simulate the disk evolution, which likely misses important evolutionary features related to the gaseous torque on the sBH. So in this work, we make use of more sophisticated 3D hydrodynamical simulations which can follow the evolution of the disk as well as the migration of the sBH and the IMBH self-consistently. Further, we use an accurate three-body dynamical code, in which general relativistic phenomena (including GW emission) are modeled in the Post Newtonian (PN) formalism \citep{2016MNRAS.461.4419B, 2021MNRAS.502.3554B}, to which we add a custom gaseous drag implementation to explore the final evolution of the sBH.


The paper is organized as follows. In Section~\ref{sec:method}, we describe the initial condition and setups of the disk, sBH and the IMBH in the hydrodynamical simulation. We try two different initial conditions of the disk, with and without the part of the disk inside the orbits of the IMBH (which are referred to as \texttt{InnerDisk} and \texttt{NoInnerDisk}), to study the viscous spread and convergence of the inner part of the disk. In Section~\ref{sec:result1}, we present the results of \texttt{InnerDisk}, to show how the migration of the sBH is accelerated by the presence of the IMBH. We also show how the evolution of the inner part of the disk affects the migration of the sBH. In Section~\ref{sec:result2} we present the results of  \texttt{NoInnerDisk}, in which initially there is no inner disk, to show that the sBH can still migrate synchronously with the IMBH even if the inner disk has been partially accreted. Based on the above results, in Section~\ref{sec:result3} we use the custom three-body dynamical simulation code to explore the final fate of the sBH. We summarize and discuss our results in Section~\ref{sec:conclusion}.

\section{Method}

\label{sec:method}

According to the analysis in Paper I (e.g. see section 2 of Paper I), an IMBH can encounter a sBH inside its orbit due to differential migration, which is then followed by the synchronized migration. We use the term synchronized migration to identify the situation where the ratio between the migration rate and the semi-major axis $\dot{a}/a$ is the same for both the IMBH and the sBH, which implies that the relative change of semi-major axis $a/a(t=0)-1$ remains the same for the two black holes. In this process, the orbital radius of the sBH is kept close to but slightly smaller than the orbital radius of the IMBH. We therefore start our simulations assuming the sBH orbit to be close to the outer IMBH with the aim of studying how the subsequent migration of the sBH is affected by the IMBH and whether synchronized migration is also seen in a more realistic 3D treatment of the problem.

Similarly to the Paper I, we will divide the whole evolution into two stages, depending whether the IMBH is close enough to the central SMBH so that GW radiation is strong enough to dominate the orbital decay. In the first stage, where the GW radiation is still negligible, we simulate the co-evolution of the disk together with the migration of sBH and IMBH, by means of 3D smooth particle hydrodynamical (SPH) code PHANTOM \citep{phantom2018}. In the simulations, the disk is modelled with millions of gas particles while the SMBH, IMBH and the sBH are modelled as three sink particles \citep{bate1995}. The SMBH is put at the center, while the sBH and the IMBH are placed on two circular orbits with similar semi-major axis. We will introduce the detailed setups and initial conditions of the hydrodynamical simulations in the next subsections. 


In the second stage, where GW radiation starts to dominate the orbital decay, we need accurate three-body dynamical simulations to model the PN evolution of the system and the possibly chaotic motion of the sBH, while introducing the effect of the disk as an effective external dissipative force. The detailed setup for these simulations will be given in the Section~\ref{sec:result3}.

\subsection{Hydro simulations initial conditions}
\label{sec:method_1}

In the hydrodynamical simulations, we choose two kinds of initial conditions for the gas disk, referred to as \texttt{InnerDisk} and \texttt{NoInnerDisk} hereinafter. The two initial conditions mainly differ at the 'inner disk' part, i.e. the part of the disk inside the orbit of the IMBH. The \texttt{InnerDisk} simulations start with a smooth disk, which extends from close to the SMBH to beyond the orbital radius of the IMBH. Conversely, in the \texttt{NoInnerDisk} simulations the gas initially only exists outside the orbit of the IMBH. We refer to this disk with the term 'outer disk' hereinafter. Some of the gas is able to flow through the gap carved by the IMBH and form an inner disk. We choose to also simulate this kind of initial condition because we found that in \texttt{InnerDisk}, the surface density of the inner disk gradually decreases because the gas that is accreted by the SMBH is only partially replenished by gas flowing across the IMBH orbit from the outer disk. 
Since the sBH is located within the inner disk, its migration may change due to decrease of the inner disk surface density. In order to assess the importance of the inner disk in the migration of the sBH we therefore perform two simulations with only the outer disk in order to set a lower limit on the inner disk surface density, which is determined by the amount of material that is able to flow through the IMBH orbit reaching the inner regions.

In both \texttt{InnerDisk} and \texttt{NoInnerDisk} cases, the initial surface density is described with the power law $\Sigma = \Sigma_0 (r/r_0)^{-p}$, where $r$ is the orbital radius and $r_0=1$ is the distance unit in the simulation and we change $p$ according to the different setup we use, assuming $p=1$ for the \texttt{InnerDisk} and $p=1.5$ for the \texttt{NoInnerDisk} case. The physical distance corresponding to $r_0$ is $100 \, R_{\rm{S}}$. The unit of time $t$ in the simulations is $P_0$, i.e. the Keplerian orbital period at $r_0$. We run two different simulations with an initial inner disc \texttt{InnerDisk}, one with $\Sigma_0=10^{-4} $ and the other one using  $\Sigma_0=10^{-5} $ in units of $M_{\rm{SMBH}} / r_0^2$, where $M_{\rm{SMBH}}$ is mass of the SMBH. We refer to these two cases with \texttt{InnerDisk$\_\Sigma$4} and \texttt{InnerDisk$\_\Sigma$5} respectively. In \texttt{NoInnerDisk}, we only use $\Sigma_0=10^{-5} $, since we want to simulate the condition when the inner disk is largely accreted as a 'lower limit' for its surface density. We choose $p=1.5$ in the \texttt{NoInnerDisk} case since the outer disk mass is more concentrated near the the orbit of the IMBH, effectively making the inner disk formation faster. 
The exact value of $p$ does not affect the final result qualitatively since the mass of the inner disk is mainly determined by the viscous evolution. 
The disk extends from 0.2 to 5.0 times the IMBH orbital radius in \texttt{InnerDisk} and from 1.0 to 2.5 in \texttt{NoInnerDisk}. Again, we choose a smaller extension of the outer disk in \texttt{NoInnerDisk} to focus on the evolution of the inner disk.

The equation of state for the disk is locally isothermal, i.e. the temperature (as well as the sound speed) of the gas is constant with time at a specific location. The sound speed $c_{\rm{s}}$ is also described by the power law $c_{\rm{s}}=c_{\rm{s,0}} (r/r_0)^{-0.5}$, which means the scale height $h/r$ is constant over the disk radial extent. We choose $c_{\rm{s,0}}$ so that $h/r = 0.02$, to simulate a relatively thin AGN disk. The viscosity of the gas is described by the Navier-Stokes viscosity model \citep{phantom2018}, where we take the shear kinematic viscosity to be proportional to the sound speed, the scale height through the viscous parameter $\alpha$ \citep{Shakura1973}. We choose the viscous parameter of the disk to be $\alpha = 0.02$ in all simulations, in order to ensure that the IMBH can successfully open a gap in the disk. We increased the number of particles until we were able to match the accretion rate onto a single central object computed from the simulations with the theoretical value for a steady state disk, i.e. $\dot{M}=3\pi\nu\Sigma$.

We found the minimum number of particles needed to resolve the chosen disk viscosity and aspect ratio to be $N=3\times 10^6$. We therefore use $N=3\times 10^6$ in the \texttt{InnerDisk} simulation to simulate the disk, while in the simulation without the inner disk \texttt{NoInnerDisk} we increase the number of particles to $N=10^7$ to better resolve the formation of the inner disk from the material that is able to cross the gap. The parameters and initial conditions for different simulation models are summarized in the Table~\ref{tab:hydro_parameters}.

\begin{table}[t]
\centering
\begin{tabular}{|c|c|c|c|c|c|} 
\hline 
Name & $\Sigma_0$ & $p$ & $r$ & $N$ & $a_{\rm{sBH}}$ \\ \hline 
\texttt{InnerDisk\_$\Sigma$4} & $10^{-4}$ & 1 & 0.2 - 5 & $3\times 10^6$ & 0.75 \\ \hline 
\texttt{InnerDisk\_$\Sigma$5} & $10^{-5}$ & 1 & 0.2 - 5 & $3\times 10^6$ & 0.75 \\ \hline
\texttt{NoInnerDisk\_075} & $10^{-5}$ & 1.5 & 1 - 2.5 & $1\times10^7$ & 0.75 \\ \hline 
\texttt{NoInnerDisk\_078} & $10^{-5}$ & 1.5 & 1 - 2.5 & $1\times10^7$ & 0.78 \\ \hline
\end{tabular}
\\
\caption{Initial condition and parameters for different simulation models. The first column is the name of different models. \texttt{InnerDisk\_$\Sigma$4} and \texttt{InnerDisk\_$\Sigma$5} belong to the same category of simulations \texttt{InnerDisk} but have different surface densities. \texttt{NoInnerDisk\_075} and \texttt{NoInnerDisk\_078} belong to the \texttt{NoInnerDisk} category but have different initial $r_{\rm{sBH}}$. The second column $\Sigma_0$ is the surface density at $r=1$. The third column $p$ is the power index of the initial surface density profile. The fourth column is the initial radius extension of the disk. The fifth column $N$ is the gas particle numbers in the simulations. The sixth column represents the initial orbital radius of the sBH. }
\label{tab:hydro_parameters} 
\end{table}

\subsection{Parameters of the sink particles}
\label{sec:method_2}

As mentioned above, the SMBH, the IMBH and the sBH are modelled as three sink particles with mass 1.0, $10^{-3}$ and $2\times 10^{-5}$ respectively, in units of the SMBH mass $10^6 M_\odot$. Note that the physical mass of the IMBH is 10 times larger than the mass of IMBH assumed in Paper I. This choice is dictated by the fact that it is harder to resolve accretion for a smaller IMBH mass in the simulations. 


\begin{figure*}[t]
\centering
\subfigbottomskip=1pt
\subfigure{
\includegraphics[width=0.25\textwidth]{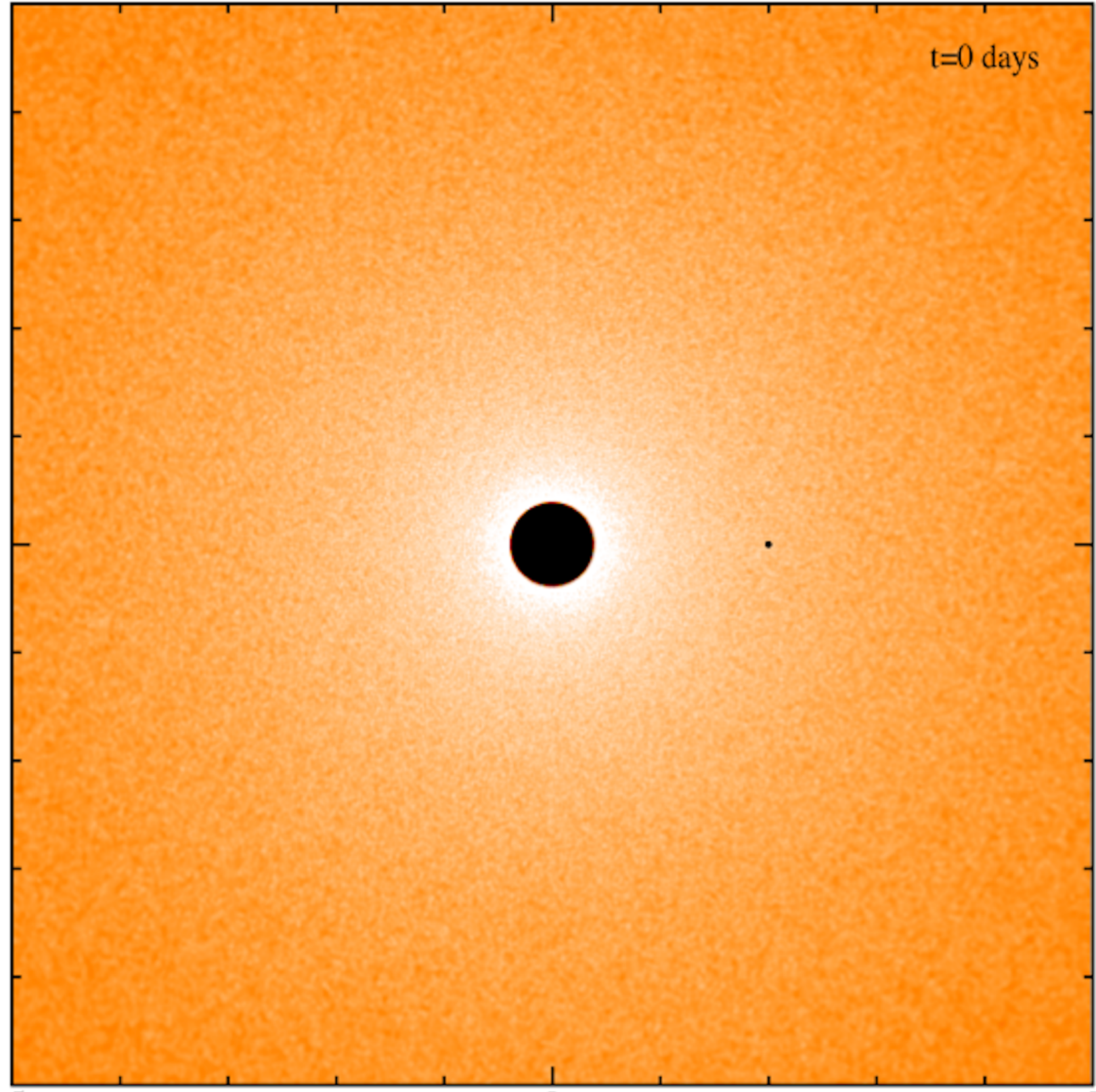}}
\quad
\subfigure{
\includegraphics[width=0.25\textwidth]{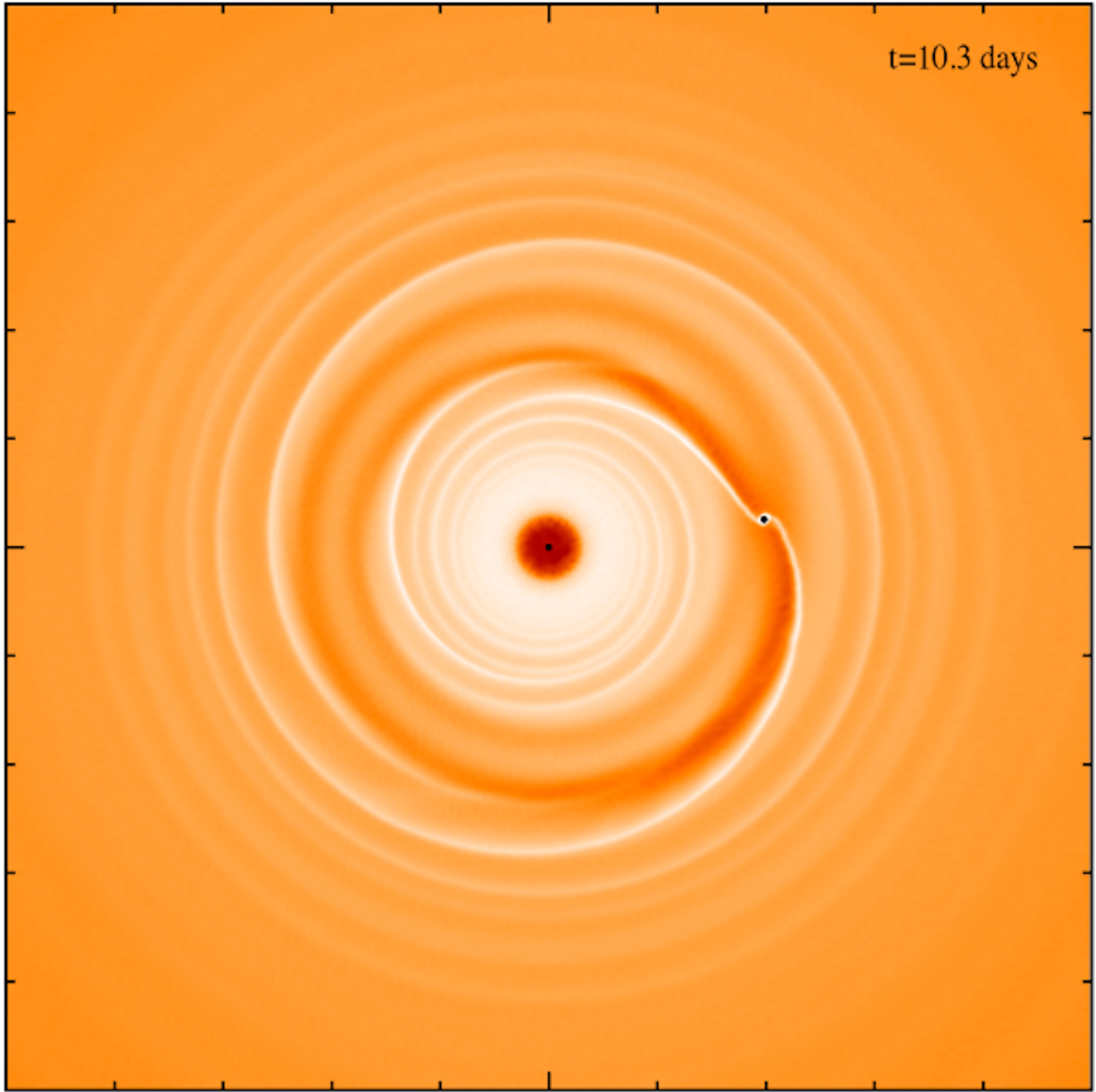}}
\quad
\subfigure{
\includegraphics[width=0.25\textwidth]{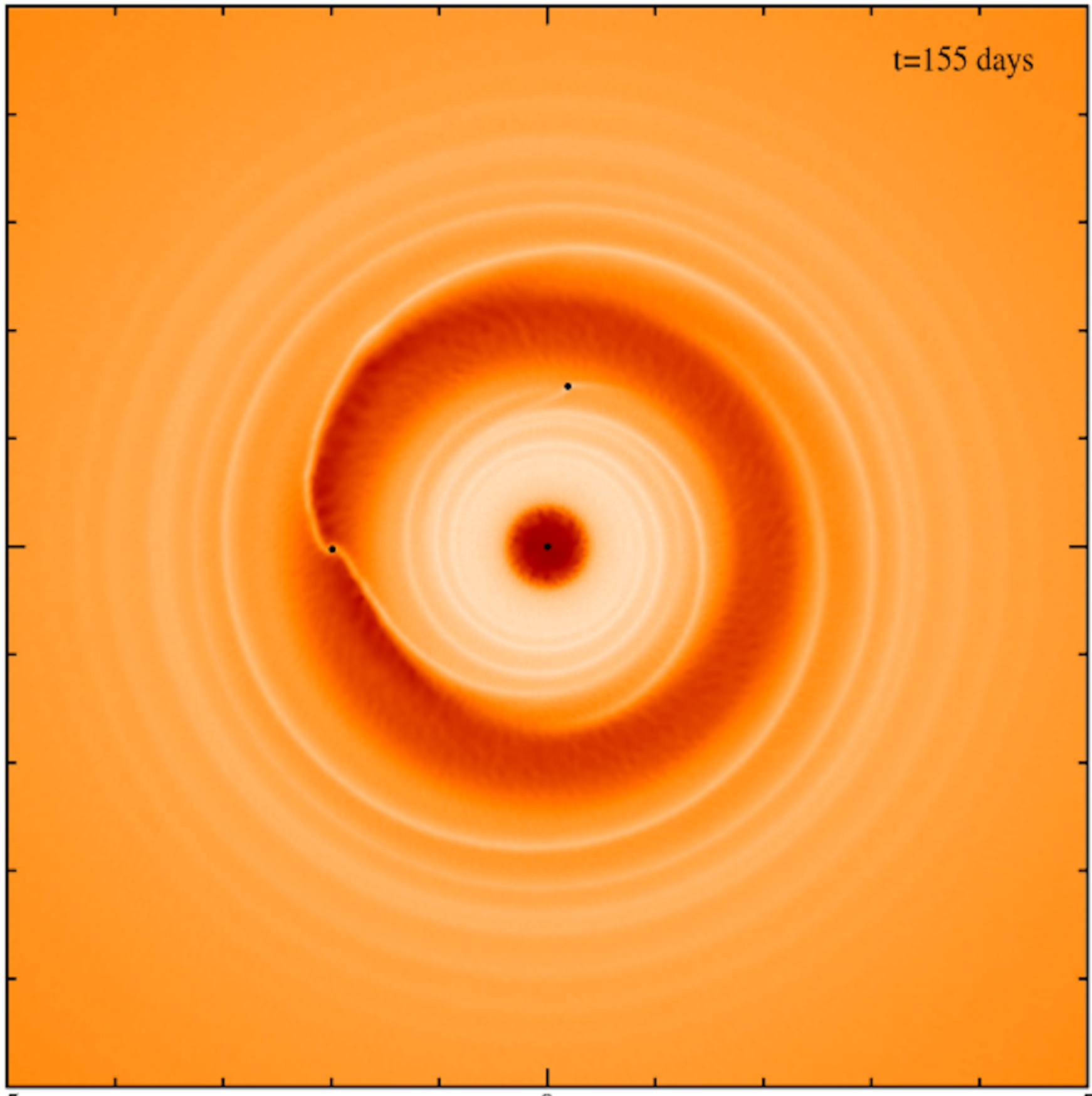}}

\subfigure{
\includegraphics[width=0.25\textwidth]{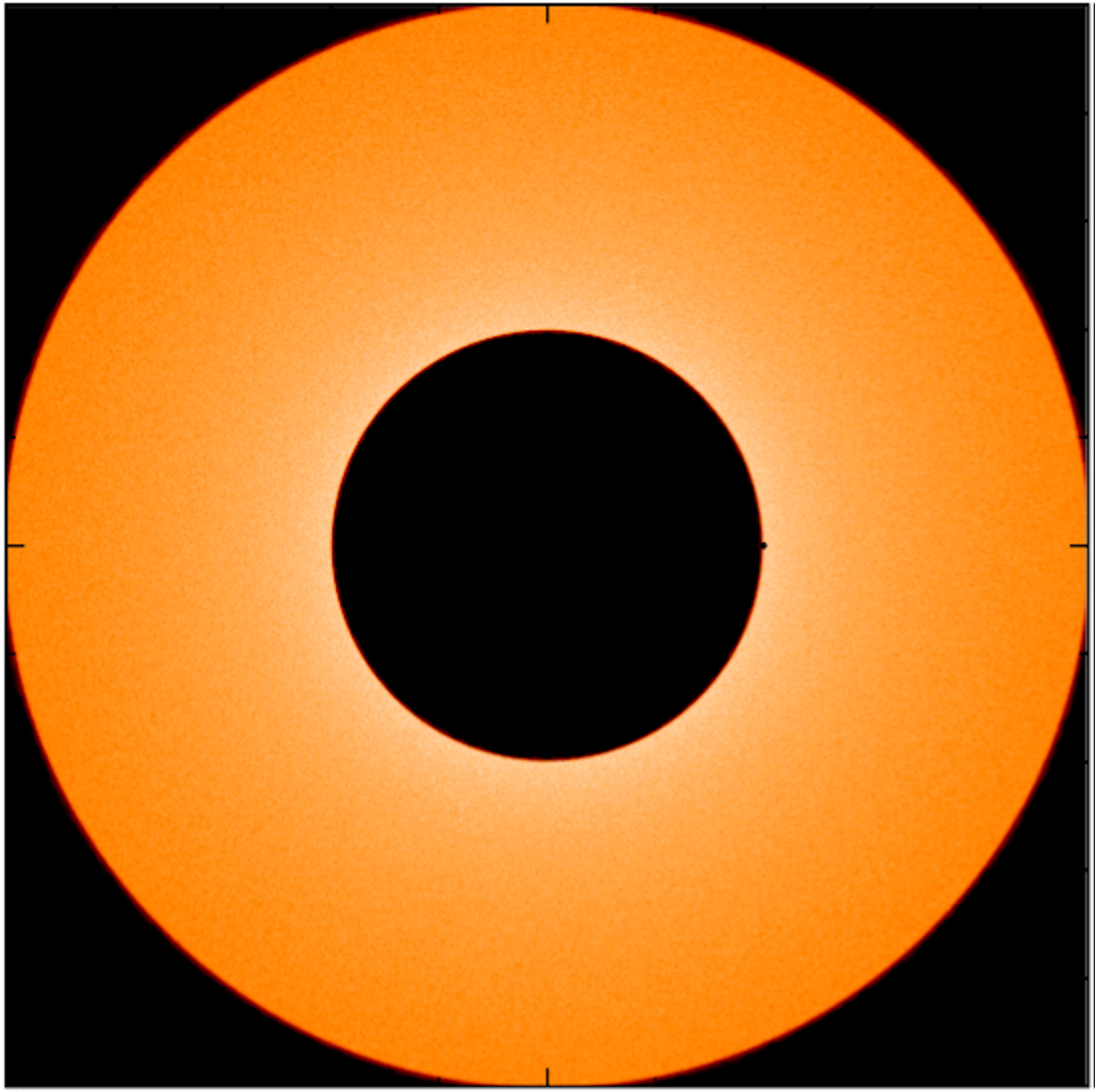}}
\quad
\subfigure{
\includegraphics[width=0.25\textwidth]{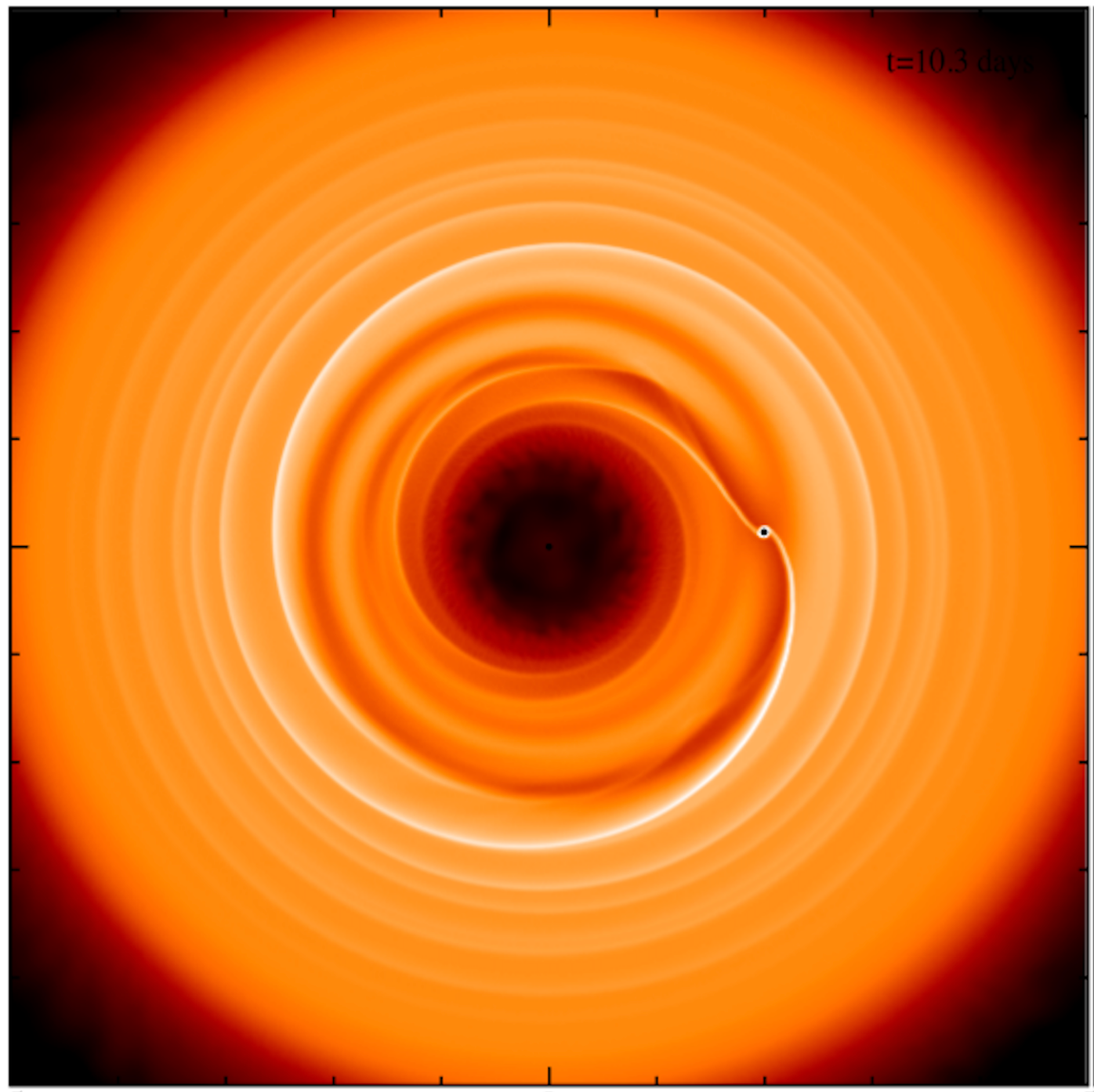}}
\quad
\subfigure{
\includegraphics[width=0.25\textwidth]{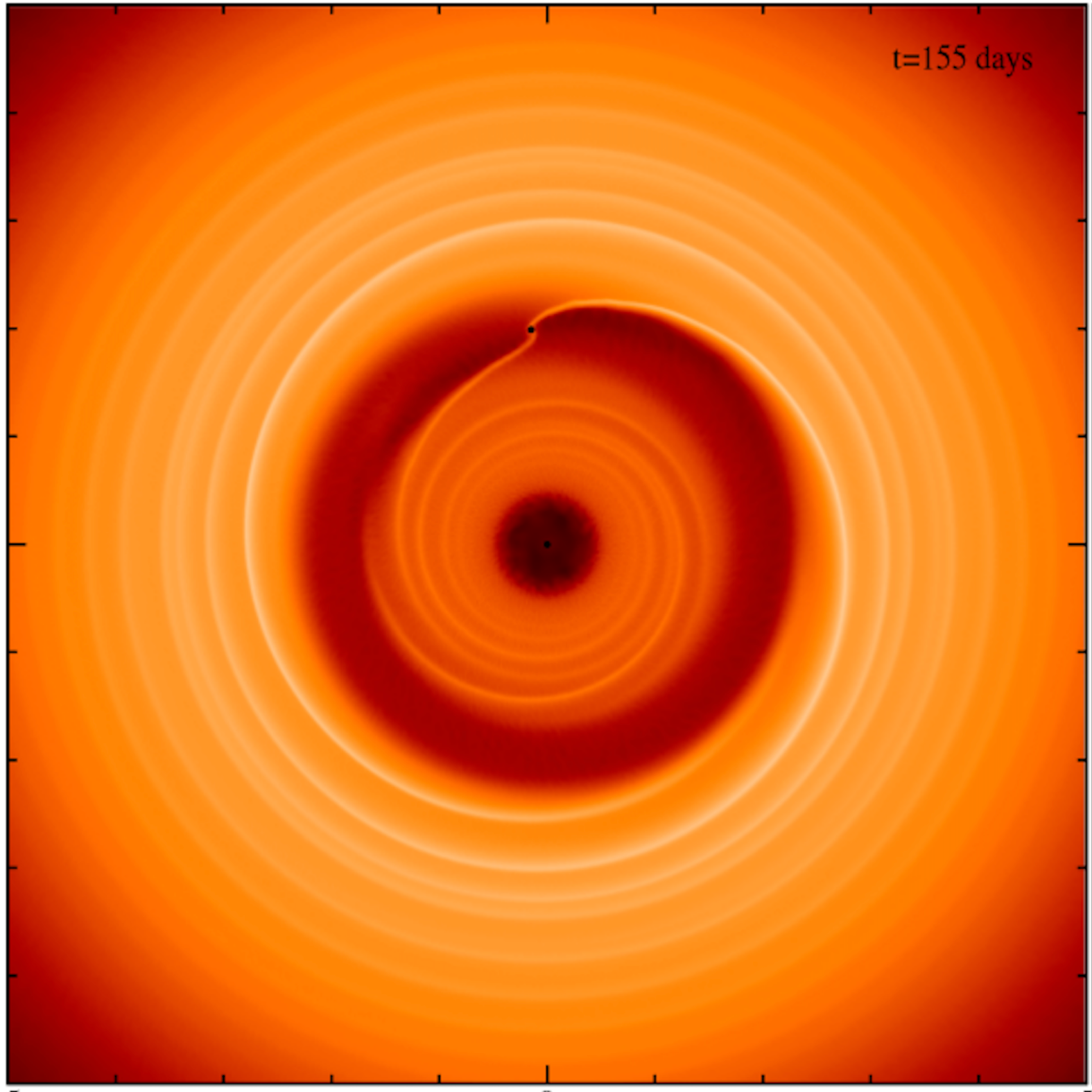}}
\caption{ Column density plots of the disk with the IMBH embedded in it. The x-axis and y-axis range from -2.5 to 2.5. Brighter color represents higher surface density and the color bar covers 5 orders of magnitude. The upper and lower panels represent the \texttt{InnerDisk\_$\Sigma$5} and \texttt{NoInnerDisk} simulation at $t = 0, \, 10, \,150 \, P_0$ respectively. }
\label{fig:snapshot}
\end{figure*}

The initial orbital semimajor axis of the IMBH is $a_{\rm{IMBH}} = r_0 = 1$, corresponding to a physical distance of $100 R_{\rm{S}}$. Since we do not take the relativistic effects into account and our disk hydrodynamics is scale free, the choice of the distance unit does not affect the results. This means that we can extend the results in Section~\ref{sec:result1} and \ref{sec:result2} to any $a_{\rm{IMBH}}$ as long as GW radiation is not important. The sBH is added into the simulation after the shape of the gap opened by the IMBH has reached a steady-state in order to avoid the effect of the transient gap-opening process on the sBH migration. The initial orbital semimajor axis of the sBH $a_{\rm{sBH}}$ is chosen to be $0.75$ for the simulations \texttt{InnerDisk}. For the simulations  \texttt{NoInnerDisk}, we choose two initial semi-major axis $a_{\rm{sBH}}=0.75, 0.78$ to study the effect of the initial location on the possibility of synchronous migration between the IMBH and sBH. We refer to this two different cases with \texttt{NoInnerDisk\_075} and \texttt{NoInnerDisk\_078}. The initial location of the sBH is already close to the edge of the gap (the width of gap is around 0.2 in simulation units). The initial orbital eccentricity of the IMBH $e_{\rm{IMBH}}$ and sBH $e_{\rm{sBH}}$ is zero. 

The accretion radius of the SMBH is set to 0.03, corresponding to $3 \, R_{\rm{S}}$. The accretion radius of the IMBH is 0.25 times its Hills radius, which is around 0.025 in simulation units. 
The IMBH mass does not increase significantly during the simulation with this accretion radius. For the sBH, we assume that its mass does not change during the simulations, since we do not resolve the accretion onto the sBH as it would require an extremely small sink radius and therefore a prohibitively high number of particles.

The gravitational force between sinks and gas particles is softened when gas particles get too close to the sinks. For both the SMBH and the IMBH, the softening radius is set to the accretion radius. The sink particles orbit will change with time as a result of the interaction with the gas particles. Therefore, the migration of the sBH and the IMBH is simulated self-consistently. We performed a test with only the sBH and the central SMBH and these simulations can recover the theoretical Type-I migration rate \citep{1980ApJ...241..425G, 1997Icar..126..261W}. 

\section{Accelerated migration and disk evolution} \label{sec:result1}

\begin{figure}[t]
\centering
\includegraphics[width=0.49\textwidth]{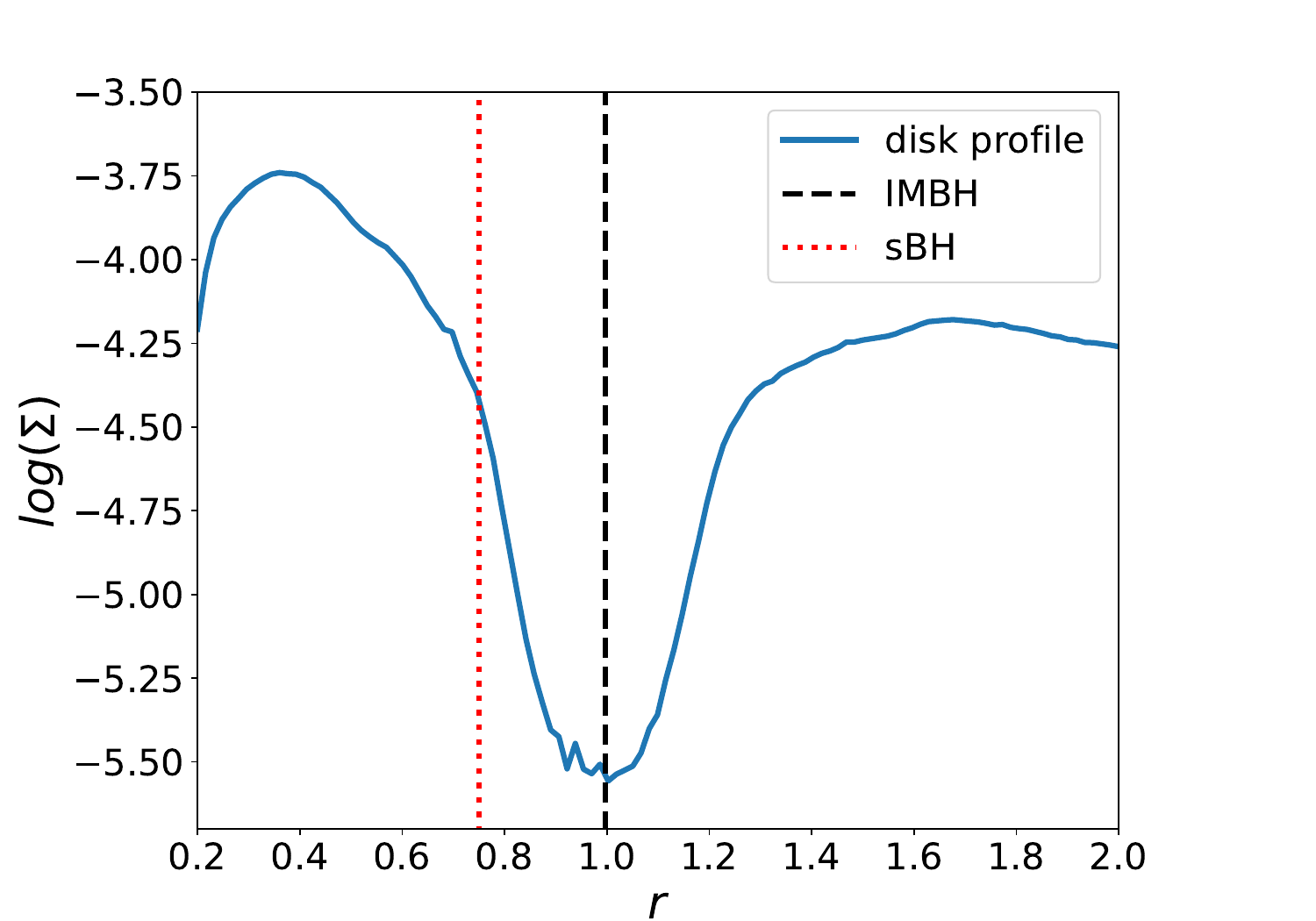}
\caption{ Disk surface density profile when the gap structure converges (after $100 \, P_0$). The x-axis and y-axis show the radius in unit of $r_0$ and the surface density of the disk in logarithmic coordinate. The locations of the sBH and the IMBH are marked with the red dotted line and the black dashed line, respectively. } \label{fig:gap} 
\end{figure}

In this section, we show the results of the simulations belonging to the \texttt{InnerDisk} set: \texttt{InnerDisk\_$\Sigma$4} and \texttt{InnerDisk\_$\Sigma$5}. The upper panel of Figure~\ref{fig:snapshot} shows the column density plots of the \texttt{InnerDisk\_$\Sigma$5} simulations at $0 \, P_0$, $10 \, P_0$ and $150 \, P_0$ respectively.
We can see the IMBH gradually carves a gap around the orbit. As mentioned in Section~\ref{sec:method}, we first let the IMBH open a gap in the smooth disk before inserting the sBH in the simulation. It takes around $100 \, P_0$ for the gap structure to reach a quasi-steady shape, at which point the gap has a depth of $\sim1.5$ dex, as shown in Figure~\ref{fig:gap}. We then place the sBH on a circular orbit very close to the IMBH and near the inner edge of the gap, as shown by the red dotted line in Figure~\ref{fig:gap}. We chose the initial separation for the sBH to be $a_{\rm sBH}=0.75$ since a sBH at a smaller separation would just evolve under type-I migration while a sBH that lies even closer to the IMBH would experience a strong tidal interaction, which would eventually halt or significantly slow down its inward migration.

We first show the migration rate of the sBH and analyze its driving mechanisms. For comparison, we also simulate the migration of a single sBH in the same initial location, but in an unperturbed disk, i.e., in the absence of the IMBH. Then we extend the simulation to a slightly longer timescale, to see whether the disk evolves significantly and whether this affects the sBH migration. 

\begin{figure}[t]
\centering
\includegraphics[width=0.49\textwidth]{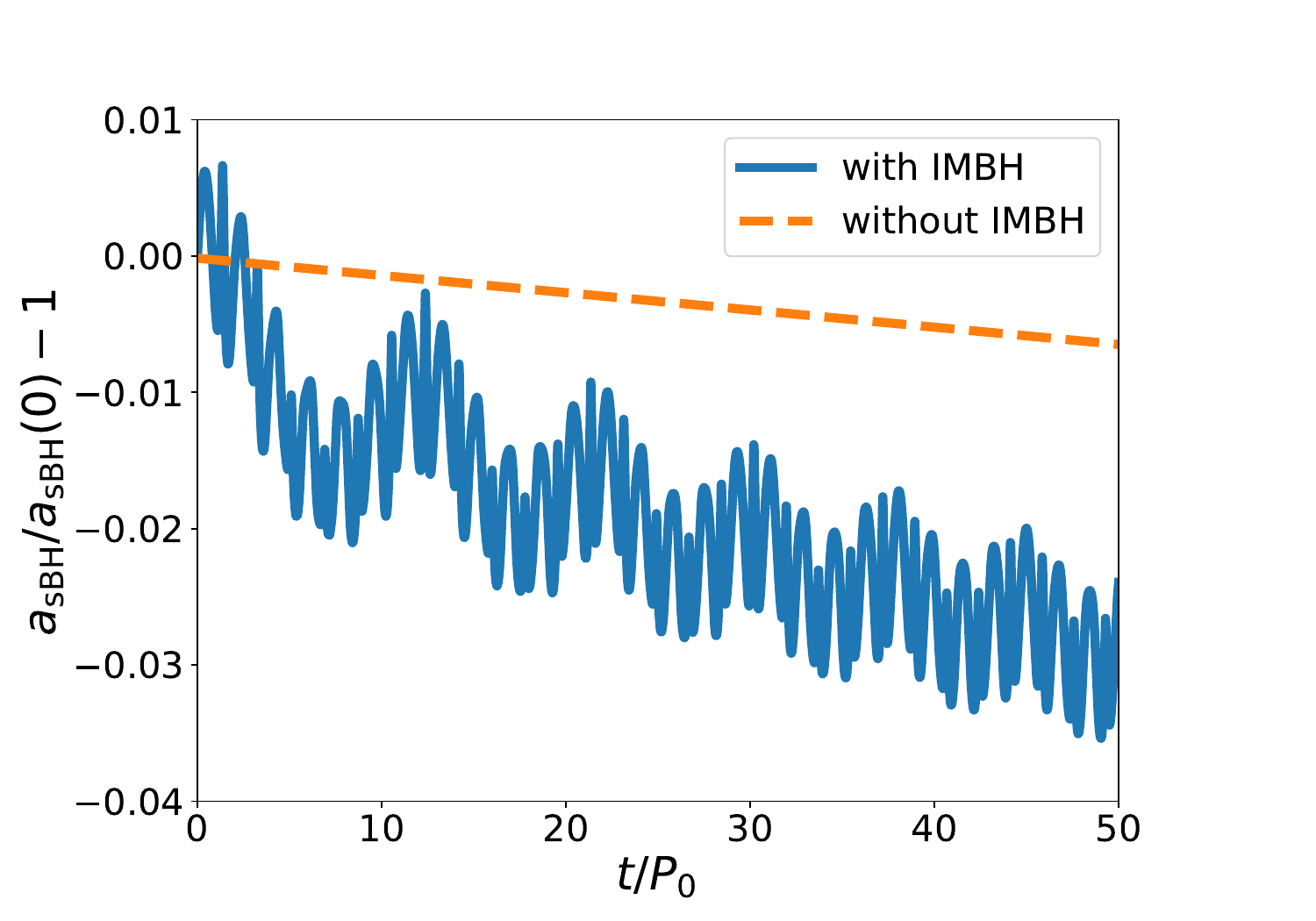}
\caption{The relative change of the orbital semimajor axis with time for the sBHs, with (blue solid line) and without (orange dashed line) the IMBH in the simulation, for \texttt{InnerDisk\_$\Sigma$4}. The x-axis represents time in units of $P_0$. The y-axis shows the change of $a_{\rm{sBH}}$ relative to the initial value. } \label{fig:acc_mig} 
\end{figure}

The sBH migration rate in \texttt{InnerDisk\_$\Sigma$4} is shown in  Figure~\ref{fig:acc_mig}, where the blue (orange) solid line represents the migration in presence (absence) of the IMBH. By comparing the overall decrease of the blue solid line with the orange dashed line, we can see that the migration of the sBH is $\sim 4 $ times faster when the IMBH is also present. So the existence of the IMBH accelerates the inward migration of the sBH, qualitatively consistent with what we have found in Paper I. We can also see that in presence of an IMBH the semi-major axis of the sBH does not decrease monotonically, but instead oscillates with time as a consequence of the gravitational tidal pull of the IMBH.

In the following, we will briefly describe the behavior of the mechanisms involved in the sBH migration: tidal torque, strengthened Type-I torque and interfering density wave. Then we will analyze the importance of these mechanisms according to the results. Additional details and equations about how these driving mechanisms affect the migration, can be found in Paper I and \citet{YangLi2024}. 

The tidal torque from the IMBH oscillates with time, as the orbital phase angles of the sBH and the IMBH change. The time averaged effect is a net sBH migration only when the two orbital periods $P_{\rm sBH}$ and $P_{\rm IMBH}$ are in one of the mean motion resonances, i.e. are a simple integer ratio of each other  \citep{1999ssd..book.....M}. When the orbit of the sBH lies at the mean motion resonance with the IMBH orbit, the average tidal effect tends to keep the integer ratio of the orbital timescales (i.e. the condition of mean motion resonance), and to increase the orbital eccentricity of the sBH. Therefore, an IMBH with a faster migration rate tends to speed up the sBH migration in order to keep the orbital timescale ratio. The strength of this effect is proportional to the mass of the IMBH, since the tidal force is proportional to the IMBH mass.

The second effect to consider is the Type-I torque from the back-reaction of the gas density wave excited by the sBH \citep{1980ApJ...241..425G, 1997Icar..126..261W}. Its strength essentially depends on the detailed profile of the disk, in particular on whether there are some asymmetries in the flow around the sBH \citep{Paardekooper11}. So the torque can be stronger when the sBH is located around the edge of the gap, where the surface density is a steep function of the radius. In general, this kind of gas torque drives the sBH to migrate inward, damps the orbital eccentricity and does not oscillate with time. The strength of this effect is proportional to the mass of the sBH and the surface density of the disk, since a heavier sBH or a more massive disk will induce a stronger density wave.

Besides the strengthened Type-I migration torque, the sBH can also be affected by the density wave excited by the IMBH. Indeed, a recent study about protoplanetary disks found that when there are two planets inside the disk, their migration is shaped not only by their own density waves, but also by the density waves excited by the companion \citep{YangLi2024}. This mechanism is known as interfering density wave, and its detailed behavior is rather complex. It can drive the sBH either to migrate inward or outward, depending on the detailed dynamics and disk parameters. When driving the sBH inward, it also damps its orbital eccentricity. Since the interfering density wave is excited by the IMBH and acts on the sBH, its behavior is somewhat similar to the tidal torque, i.e. the strength depends on the IMBH mass and oscillates with time. On the other hand, since it is a gaseous torque, it also depends on the surface density, similar to the Type-I torque.

\begin{figure}[t]
\centering
\includegraphics[width=0.45\textwidth]{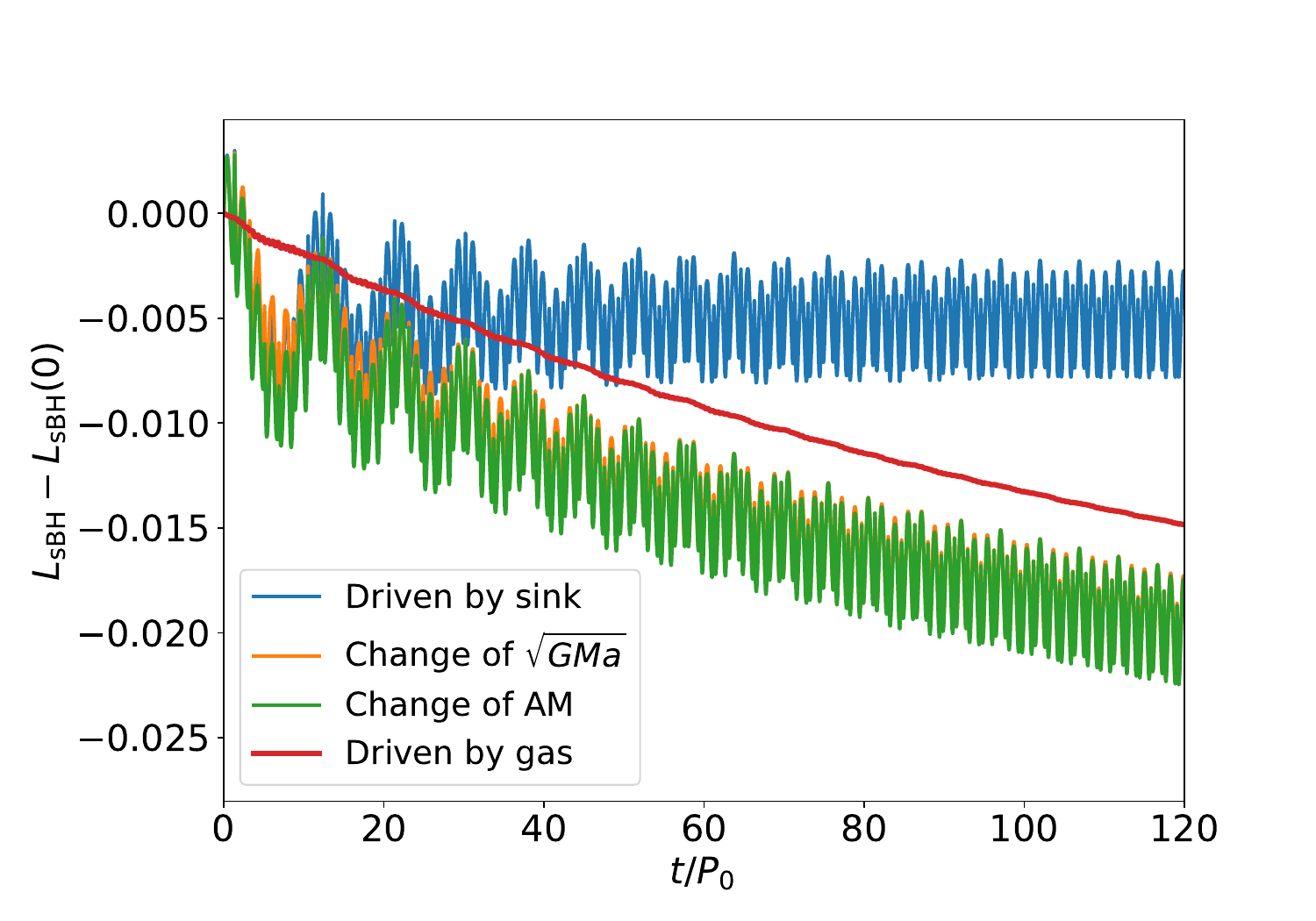}
\caption{ Integrated torques and specific orbital angular momentum change of the sBH for \texttt{InnerDisk\_$\Sigma$4}. The red and the blue line represent the time integrated torque exerted by gas particles and by the IMBH, respectively. The green line represents the total evolution of the orbital angular momentum of the sBH, while the orange line represents the contribution of the semimajor axis evolution on the angular momentum change. The x-axis represents the time in the unit of $P_0$. The y-axis shows how large the specific angular momentum has changed, or how large the torque has contributed. }
\label{fig:gas_mig_largeS}
\end{figure}

\begin{figure}[t]
\centering
\includegraphics[width=0.45\textwidth]{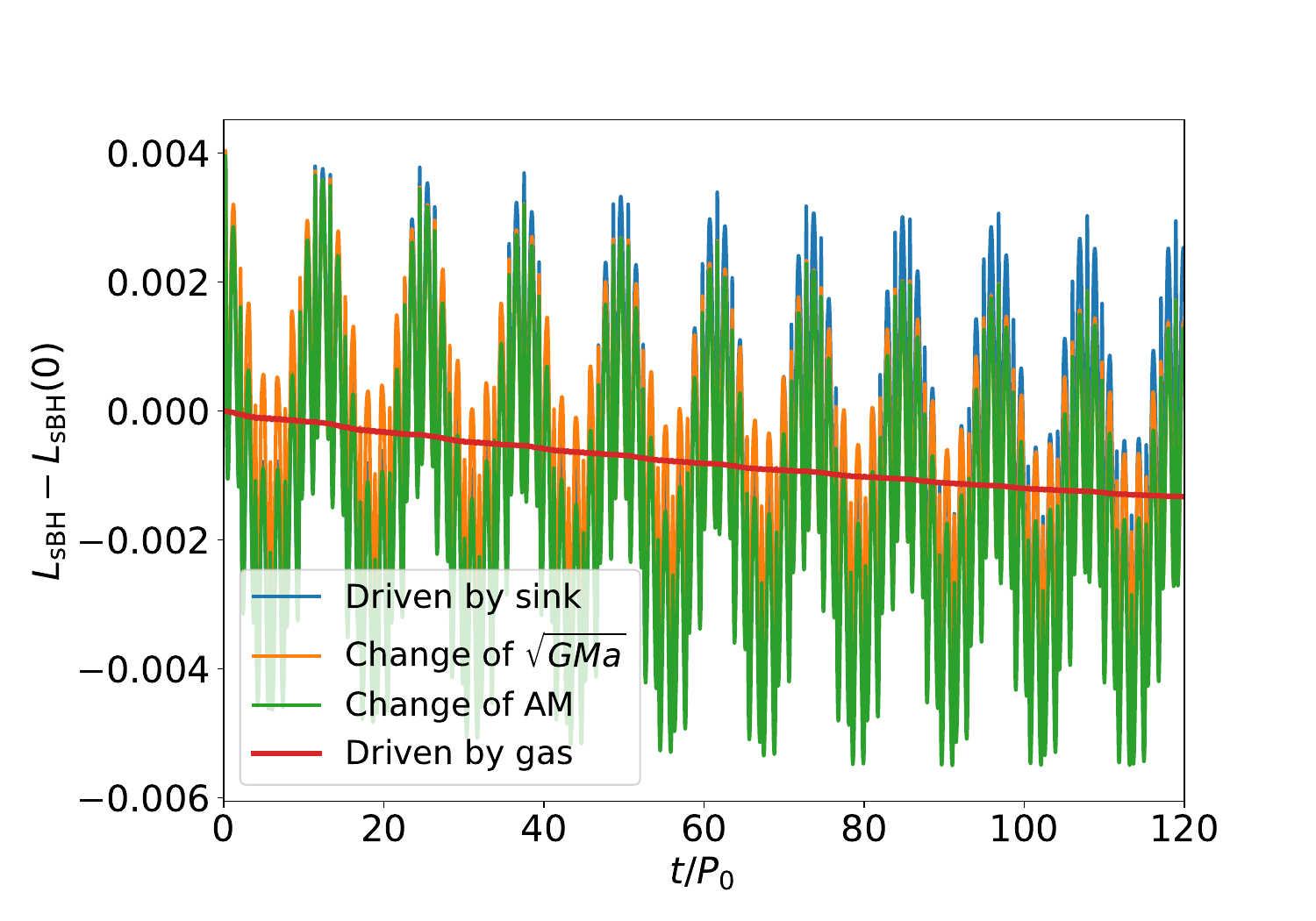}
\caption{ Same as Figure~\ref{fig:gas_mig_largeS}, but for \texttt{InnerDisk\_$\Sigma$5}} \label{fig:gas_mig_smallS} 
\end{figure}

To understand whether it is the tidal torque directly from the IMBH or the two gaseous effects that accelerate the sBH migration, we separate the force acting on the sBH in the simulation in two contributions: the force from the gas particles and the force from the sink particles. We integrate in time the torques that the gas particles and the IMBH exert on the sBH separately, and compare them with the orbital angular momentum change of the sBH during the simulation. We note that, since the change of the specific orbital angular momentum $L_{\rm{sBH}} \propto a_{\rm{sBH}}^{1/2} (1-e_{\rm{sBH}}^2)^{1/2}$ also includes the effect of the orbital eccentricity change, we also calculate the evolution of $a_{\rm{sBH}}^{1/2}$ separately and compare it with the above terms. We perform the same calculations for both our chosen values of disk surface density, i.e. \texttt{InnerDisk\_$\Sigma$4} and \texttt{InnerDisk\_$\Sigma$5}. The results are shown in the Figure~\ref{fig:gas_mig_largeS} and \ref{fig:gas_mig_smallS}, respectively.

In Figure~\ref{fig:gas_mig_largeS}, we can see that the orange and the green line, representing the contribution of the semi-major axis change to the angular momentum and the evolution of the angular momentum itself respectively, vary almost identically over time. That means that the change in angular momentum is mostly due to the migration of the sBH (evolution of $a_{\rm{sBH}}$), rather than the eccentricity change. The small discrepancy represents the contribution of eccentricity evolution on the angular momentum. This discrepancy decreases with time as the orbital eccentricity is damped by the gas torque. We can then compare the green line with the red and blue lines, to see the contribution of tidal torque and the gaseous torque on the migration. The angular momentum is wiggling while decreasing, similarly to the results in Figure~\ref{fig:acc_mig}. The decreasing trend matches quite well the red line, which represents the amount of the angular momentum dissipated by the gas. This naturally implies that most of the angular momentum is extracted by gaseous torques. The wiggling part of the green line matches in amplitude and phase the blue line, meaning that the tidal torque dominates the evolution of $a_{\rm{sBH}}$ on short timescales, i.e. within the first $\sim 10 \, P_0$. Compared to the red line, the overall decrease of the blue line is much smaller, which means the accumulated angular momentum change driven by the tidal torque is much smaller than that driven by the gaseous effect. We can therefore conclude that in \texttt{InnerDisk\_$\Sigma$4}, the gaseous torque drives the accelerated migration of the sBH.

In Figure~\ref{fig:gas_mig_smallS}, we show the results of the same analysis performed on the \texttt{InnerDisk\_$\Sigma$5} simulation, in which there is less gas and thus the gaseous torque is weaker. We can see that, compared to \texttt{InnerDisk\_$\Sigma$4}, the overall angular momentum decreasing rate (green line), as well as the contribution of gaseous effect the the migration (red line), are smaller (notice the different $y$-axis scale compared to Figure~\ref{fig:gas_mig_largeS}). The reason is that the surface density is 10 times smaller in this case, so the migration rate of the IMBH and the sBH, driven by the gaseous effect, is also roughly 10 times smaller. Furthermore, the smaller $\Sigma_0$ also results in weaker eccentricity damping by the gas. So the discrepancy between the green and orange line, which represents the contribution of the eccentricity on the angular momentum evolution, hardly decreased in $\sim 100 \, P_0$. Relatively, the overall decrease of the blue line is more significant, which means the contribution of tidal torque on the migration is more important in this case but still comparable with the gas induced migration. 

The results of the two simulations demonstrates that both the gaseous torque and the tidal torque can drive the accelerated migration. The gas torque dominates in case of high disk surface densities, while tidal torque plays a major role when the disk surface density is low. We further analyze the contribution of the two types of the gaseous torque, i.e. the strengthened Type-I torque and the interfering density wave, on the accelerated migration. We found that the density wave excited by the IMBH dominates the evolution of the sBH orbit, at least initially. But as the orbital eccentricity is damped by the gas, the strength of the interfering density wave will decrease so that the strengthed Type-I torque will dominate.

\begin{figure}[t]
\centering
\includegraphics[width=0.49\textwidth]{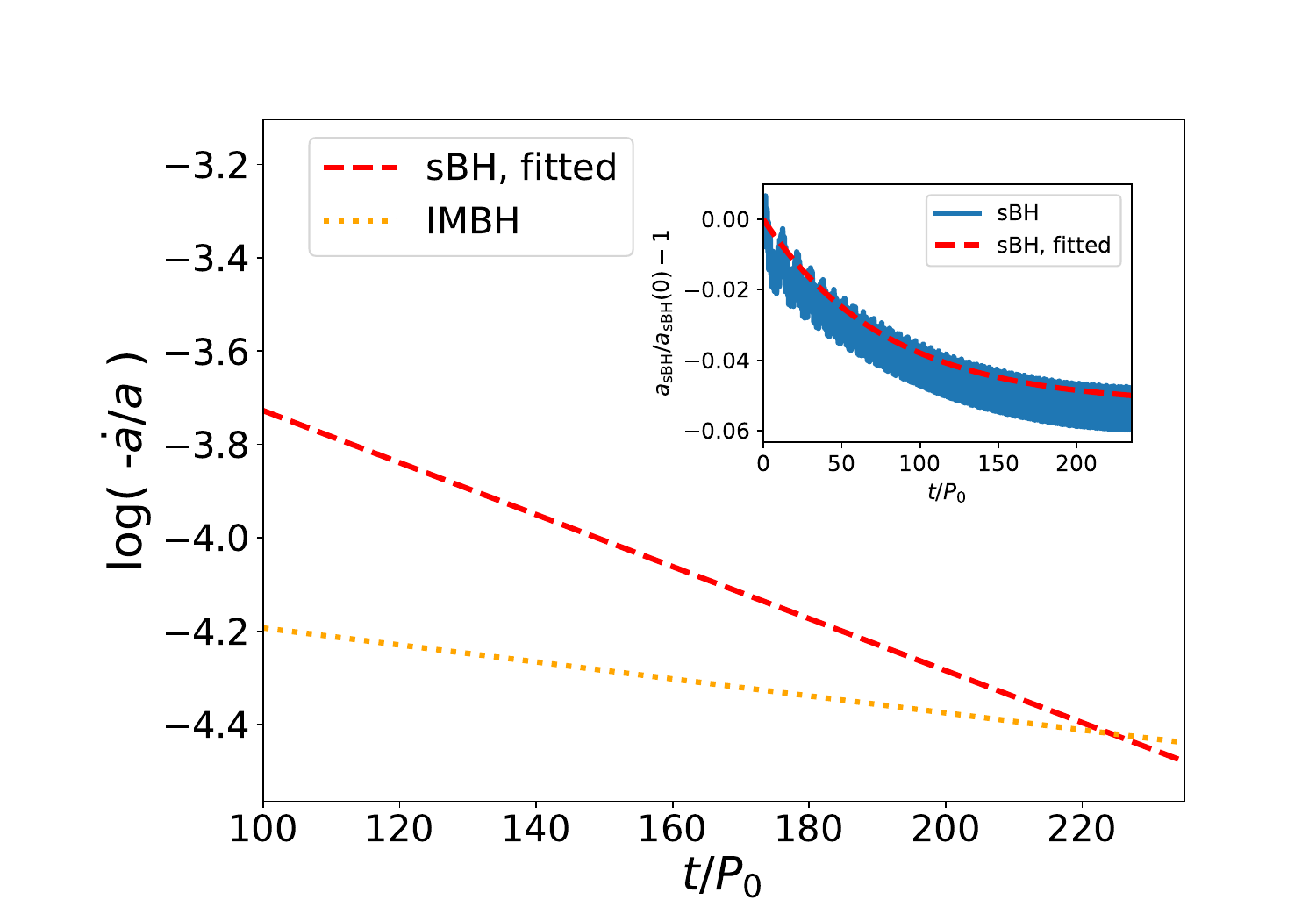}
\caption{ Migration rate evolution of the sBH and IMBH. For clarity, we fit the semimajor axis evolution of the sBH with a simple exponential function, which can capture the general decreasing trend, shown by the inset.
In the main figure, the red dashed line and the orange dotted line represent the migration rate of the sBH and the IMBH respectively. In the inset, the blue solid line represents the raw evolution of $a_{\rm{sBH}}$, while the red dashed line represents the fitted decreasing trend of $a_{\rm{sBH}}$. The x-axis shows the time in unit of $P_0$. The y-axis of the main figure shows the migration rate normalized with semimajor axis, in logarithmic coordinate. The y-axis of inset shows the relative change of $a_{\rm{sBH}}$. } \label{fig:mig_evolve} 
\end{figure}

In Figure~\ref{fig:mig_evolve} we show the long term migration rate of the sBH compared to that of the IMBH in \texttt{InnerDisk\_$\Sigma$4}. To show the sBH migration more clearly, we fit the average decreasing trend of $a_{\rm{sBH}}$ with an exponential function as $\dot{a}_{\rm{sBH}}/a_{\rm{sBH}} = 0.0006 \, {\rm{exp}} (-0.013 \, t/P_0) $, where $\dot{a}_{\rm{sBH}}$ is the change rate of sBH semi-major axis. The comparison of fitting and raw evolution is shown in the inset of Figure~\ref{fig:mig_evolve}. By comparing the red dashed and the orange dotted line, we can see that initially the migration rate of the sBH is higher. However, the migration rate of the sBH decreases significantly faster compared to that of the IMBH. Around $t = 240 \, P_0$, the migration rate of the sBH becomes smaller than the IMBH rate. The exact time at which this transition happens may be affected by the fitting of the very complex decay behaviour. A similar but less evident phenomenon also happens in the case of a lower surface density disk, i.e. in our simulation \texttt{InnerDisk\_$\Sigma$5}. 

\begin{figure}[t]
\centering
\includegraphics[width=0.48\textwidth]{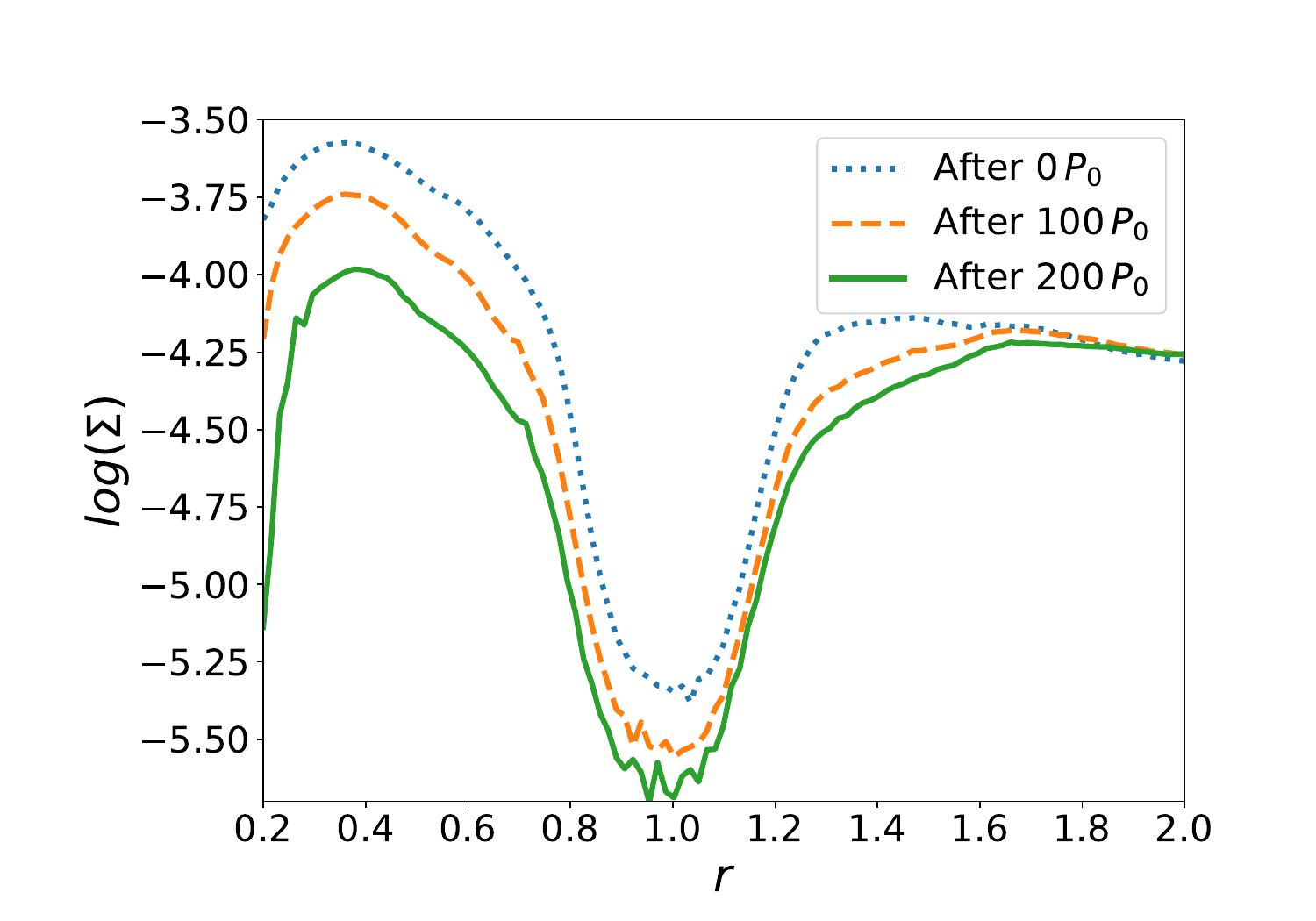}
\caption{ Disk surface density profile evolution after the sBH is put into the simulation, for \texttt{InnerDisk\_$\Sigma$4}. Blue dotted, orange dashed and the green solid lines represent the surface density profile right after, $100 \, P_0$ after, and $200 \, P_0$ after the sBH is inserted in the simulation, respectively. The x-axis shows the radius. The y-axis shows the surface density in logarithmic coordinate. } \label{fig:disk_evolve} 
\end{figure}

The above decreasing of the sBH migration rate results from the fact that the disk evolves significantly, especially the surface density of the inner disk compared to the overall disk profile. In Figure~\ref{fig:disk_evolve} we show the disk surface density profile evolution over $200 \, P_0$ in \texttt{InnerDisk\_$\Sigma$4}. Different lines represent the profiles at different times. We can see that the surface density of the inner disk decreases by about 0.5 dex, while the surface density around the gap and outer disk decreases more slowly and it is essentially constant at $r>1.8$. This is a natural result of the accretion of the gas in the inner disk on the central SMBH. In general, the surface density of the inner disk will continue to decrease although this part of the disk will be partially replenished by the gas that is able to flow through the gap carved by the IMBH. As we have shown in Figure~\ref{fig:gas_mig_largeS} and \ref{fig:gas_mig_smallS}, the inner disk surface density determines the strength of the gaseous torque, which contributes significantly to the migration of the sBH. It is however difficult to make predictions on the synchronized migration on longer timescales as the sBH migration rate is continuously decreasing due to the disk evolution. In order to place a lower limit on this migration rate, we initialize the same simulation but without the inner disk. This would allow us to measure the migration rate of the sBH due to the sole effect of the gas that is able to flow through the IMBH gap, effectively producing an inner disk. We present our results in the next section.


\section{Lower limit of inner disk and synchronized migration}

\label{sec:result2}

Since the inner disk will be gradually accreted, which will then affect the migration of the sBH, it is necessary to investigate whether there will still be enough gas in the inner disk after a long enough time (i.e. of the order of the migration timescale), to sustain the sBH migration and possibly induce a synchronous migration with the IMBH. We therefore ran two simulations without the part of the disc that lies inside the IMBH orbit in order to place a lower limit on the sBH migration rate driven by the material that is able to overcome the IMBH gap. The parameters of the simulations are shown in Table~\ref{tab:hydro_parameters}. We initialize an outer disk with initial surface density $\Sigma_0 = 10^{-5}$ since we aim at placing a lower limit on the sBH migration rate.

\begin{figure}[t]
\centering
\includegraphics[width=0.49\textwidth]{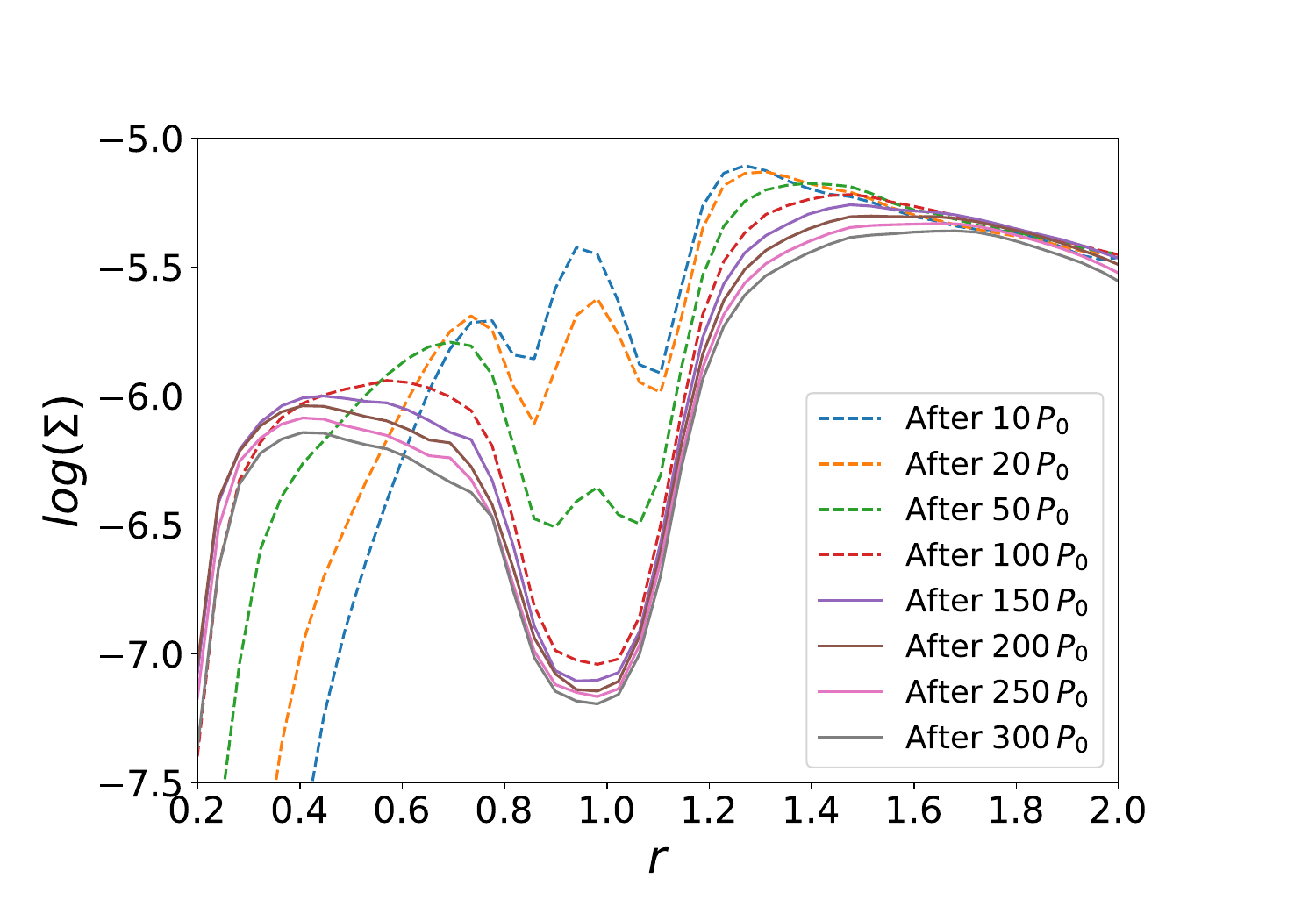}
\caption{ Disk surface density profile evolution for the result of \texttt{NoInnerDisk}. The blue, orange, green and red dashed line represents the disk profile 10, 20, 50 and 100 $P_0$ after the start of the simulation. The purple, brown, pink and grey solid line represents the disk profile 150, 200, 250 and 300 $P_0$ after the start of the simulation. The x-axis shows radius. The y-axis shows the surface density in logarithmic coordinate. } \label{fig:disk_evolve_2} 
\end{figure}

Figure~\ref{fig:disk_evolve_2} shows the evolution of the disk surface density and in particular how much gas is able to flow through the gap and feed an inner disk.
We show the disk column density for the case \texttt{NoInnerDisk} where we only include the IMBH, in the lower panels of the Figure~\ref{fig:snapshot}. The left, middle and right plots represent the snapshot at $0 \, P_0$, $10 \, P_0$ and $150 \, P_0$. 
We can see the gas in the outer disk gradually flows in and the IMBH gradually curves a gap around its orbit. 
The mass of the inner disk gradually increases and is clearly separated from the outer disk by the gap. 
After its formation, the inner disk evolves very little and the profile finally converges to a steady state after roughly $150\,P_0$. 

We note here that we do not replenish the outer disk with a steady inflow of material, i.e. our disk has a finite mass. We find the inner disk to reach a quasi-steady state equilibrium between the material coming from the outer disk and the material that is accreted onto the SMBH.
We insert the sBH into the simulation after the formation of the IMBH-carved gap, at $t=150\,P_0$.
We explored two different initial positions of the sBH, i.e. $r =0.75$ and $r = 0.78$, and we refer to them with \texttt{NoInnerDisk075} and \texttt{NoInnerDisk078}. The disk profile evolution is almost the same for these two simulations. We increase the initial sBH separation to $r = 0.78$ in order to potentially aid the synchronous migration of the two objects in the disk. We here note that the results in \texttt{NoInnerDisk075} show that the sBH still migrates more slowly than the IMBH, somewhat in contrast with the results of Paper I, since the migration rate is sensitive to the initial choice of the sBH separation. Nevertheless, when the initial separation decreases due to the slower migration of the sBH, the synchronous migration condition can be reached, as we will show in the following. 

\begin{figure}[t]
\centering
\includegraphics[width=0.49\textwidth]{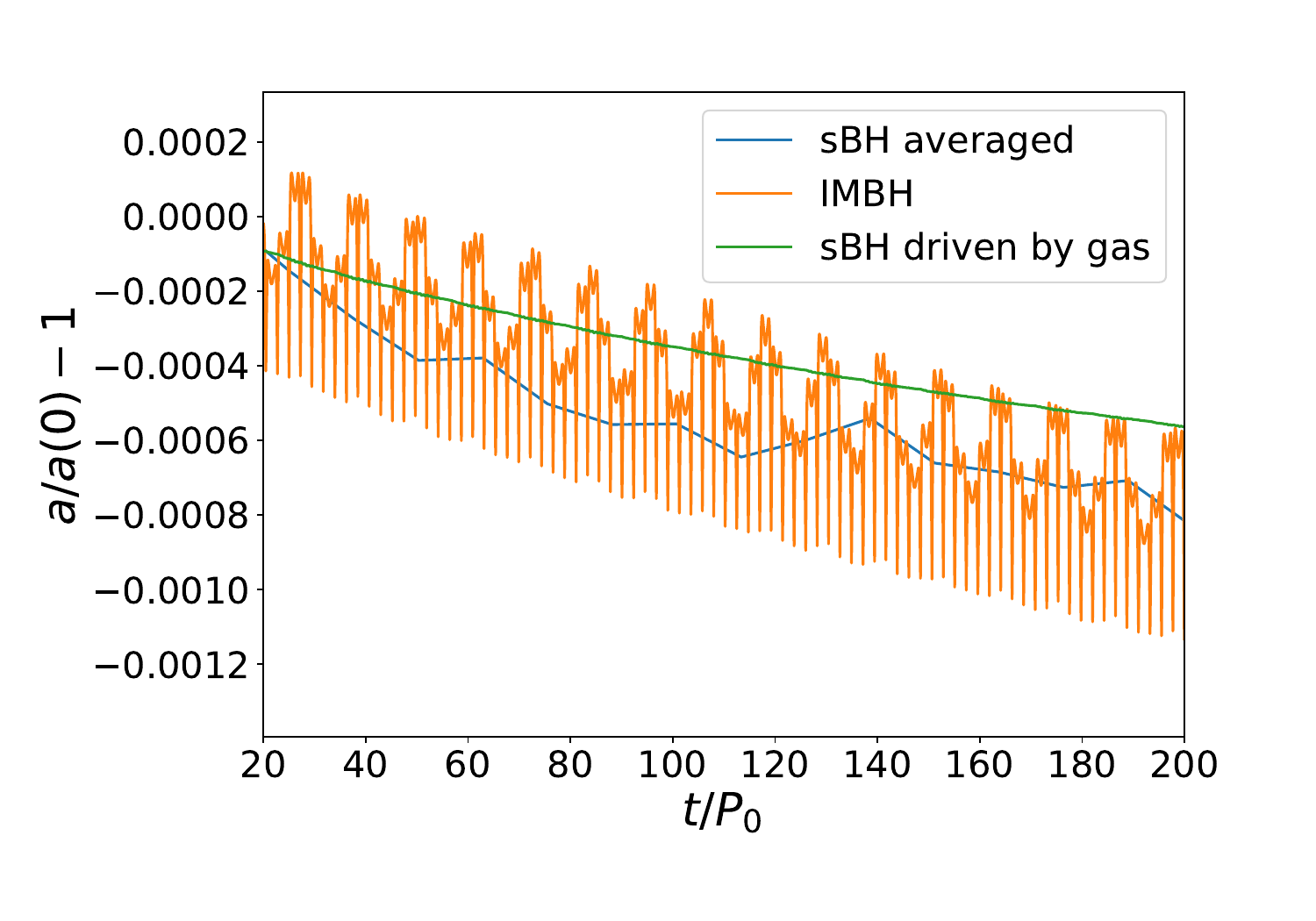}
\caption{ The migration of the sBH and IMBH in \texttt{NoInnerDisk\_078}, as well as the migration rate of the sBH contributed by the gaseous effect. The blue, green and the orange line represents the evolution of $a_{\rm{sBH}}$, the contribution due to the gas and the evolution of $a_{\rm{IMBH}}$ . The x-axis shows the time in unit of $P_0$. The y-axis shows the relative change of semi-major axis compared to the initial value. } \label{fig:mig_aver_078} 
\end{figure}

The results of \texttt{NoInnerDisk\_078}are shown in the Figure~\ref{fig:mig_aver_078}. We computed the average of the sBH migration rate over $15$ orbits to reduce the wiggling induced by the very close IMBH tidal torque. The blue and the orange lines represent the migration of the sBH after the average procedure and the migration of the IMBH respectively, while the green line represents the migration rate of the sBH induced by the gaseous torque. 
Comparing the sBH migration rate here $ \dot{a}_{\rm{sBH}} \sim -3 \times 10^{-6} \, a_{\rm{sBH}} $ to that in Figure~\ref{fig:gas_mig_smallS} $ \dot{a}_{\rm{sBH}} \sim -1 \times 10^{-5} \, a_{\rm{sBH}} $, we can see that the migration rate here is several times smaller. The reason is that is this case the inner disk is less massive than the case in Figure~\ref{fig:gas_mig_smallS}, which results in an ineffective gaseous torque. By comparing the decrease of the blue line and the orange line, we can see the migration rate of the sBH in this case is almost the same as the IMBH. So the sBH migrates synchronously with the IMBH. Furthermore, in this case, the sBH migration does not slow down a lot like found in Figure~\ref{fig:mig_evolve}, since in \texttt{NoInnerDisk} the whole disk profile converges. So we can say that in this case, synchronized migration between the sBH and the IMBH can be kept all along the migration.

In conclusion, with both the assistance of the gaseous torque and the tidal torque, the sBH can migrate synchronously with the IMBH even when the inner disk has been partially accreted.
or the parameters of our simulation, the equilibrium point at which synchronous migration occurs is $0.78 \, a_{\rm{IMBH}}$. The exact value for synchronous migration might depend on the exact profile and width of the gap as well as on the IMBH mass. 
We caution that it might be still possible to induce synchronous migration even if the initial separation between the sBH and the IMBH is larger as the IMBH would migrate faster than the sBH, therefore reducing their relative separation over time. One possible reason why we do not find synchronous migration for an initial separation of $r=0.75$ is that we would need to evolve the system for a significantly longer time and this is computationally expensive.
Note that, because of the nature of the torques driving the migration, this conclusion does not depend on the specific physical scale and can be extended to any physical radius along the migration process as long as the GW radiation does not dominate the dynamics, provided that the disk surface density profile has reached a steady state. Indeed, all the related timescales (e.g. migration timescale, viscous and accretion timescale) will scale with the same factor, which will not change the migration behavior of the sBH. Therefore, once the IMBH has encountered one sBH inside its orbit, they will migrate inward synchronously until the GW radiation starts to be important. We study this later stage of the evolution by means of custom three-body simulations in the next section. 

\section{Later evolution in GW dominated scheme}
 
\label{sec:result3}
 
The synchronized migration shown in the last section will continue so long as the gaseous and tidal torques are the dominant physical mechanisms governing the dynamics of the system. As the IMBH and sBH orbit shrink, however, GW emission will become more and more efficient and will eventually dominate their orbital evolution around the SMBH. Since the shrinking rate $\dot{a}_{\rm GW}$ is a strong function of the mass, the IMBH will start to inspiral much faster than the sBH, catching up with it.
At this point, the orbits of the sBH will be more and more chaotic, according to the tests done in Paper I. Whether the sBH will be kicked out, fall into the SMBH or captured by the IMBH, is unclear. So here we use a three-body numerical integration code, implemented with PN dynamics and a fictitious force mimicking gaseous drag, to explore the final fate of the sBH in the GW dominated phase and the related detectable phenomena.
 
\subsection{Dynamical code and gaseous force}
 
We use the code developed in \cite{2016MNRAS.461.4419B, 2021MNRAS.502.3554B}. The code integrates the equations of motion of a three-body system evolving the PN dynamics up to order 2.5PN, by directly solving the three-body trajectories from the PN Hamiltonian. This allows us to self-consistently calculate the GW radiation between the sBH and the IMBH when they get close to each other, as well as the orbital decay of the sBH and the IMBH around the SMBH. The initial orbital radius of the IMBH is set to $a_{\rm{IMBH}} = 30 \, R_{\rm{S}}$, where the orbital decay timescale due to the GW radiation becomes shorter than the migration timescale driven by the gas. The initial orbital radius of the sBH is $a_{\rm{sBH}} = 20 \, R_{\rm{S}}$, such that $a_{\rm{sBH}}/ a_{\rm{IMBH}} = 0.67$. This is slightly smaller than the value we get in the last section and this allows a relaxation stage before the synchronized migration. The initial orbital eccentricity of the sBH is $e_{\rm{sBH}} = 0.05$, and for IMBH it is $e_{\rm{IMBH}} = 0.01$, although the exact value does not affect the result qualitatively. We will keep the initial phase angle of the IMBH fixed to 0, while trying different values between 0--$2\pi$ for the sBH, to test the effect of the randomness on the results. In the following, we use $R_{\rm{S}}$ as the length unit, and 1 yr as the time unit.
 
Since there are no gas particles in the dynamical code, we need to implement the gaseous torque on the sBH and the IMBH through the introduction of a fictitious force, i.e. an extra {\it ad hoc} force added to the simulation besides the gravity among the three-body system. This fictitious force will produce the same effect of driving migration and damping eccentricity as the 'true' gaseous drag we get in the hydrodynamical simulation. So we use the gaseous torque results presented in the last two sections to calibrate the force strength and how it depends on the orbital elements. We then test the effect of this fictitious force in the dynamical code for the two-body case, i.e. simulation with only the IMBH and SMBH or the sBH and SMBH, to confirm that it reproduces the effect of the gaseous torque.
 
We keep the above gaseous force model constant for most of time in the simulation, since we assume that during the evolution covered by the dynamical simulation, the surface density of the inner disk does not change a lot. This assumption is reasonable according to the simulation result in \cite{2016MNRAS.457..939C}, which simulates how the inner disk is squeezed by the IMBH when the orbital decay of the IMBH is dominated by GW radiation. They found that the squeezing effect on the disk surface density is negligible until the IMBH reaches $r\sim 5 \, R_{\rm{S}}$. In our simulation, most of the interesting dynamical interplay between the IMBH and the sBH occurs at larger distances.
So it is reasonable to assume a constant gaseous force model in our study. Conversely, when the orbit of the sBH is so chaotic that the location is deep into the region of the gap of the IMBH, we decrease the gas force by one dex to model the low gas density in the gap, since the gas surface density inside the gap is about one dex lower than the surface density outside the gap (see Figure~\ref{fig:disk_evolve_2}). Furthermore, when the distance between the sBH and the IMBH is too small (smaller than $0.01 \, a_{\rm{IMBH}}$), we will turn off the gaseous force term too avoid any unpredictable spurious interactions, since we do not know how the gaseous force will affect the evolution when the motion of the sBH is dominated by the IMBH gravity.

\subsection{Outcome 1: swap and ejection}

\begin{figure}[t]
\centering
\includegraphics[width=0.49\textwidth]{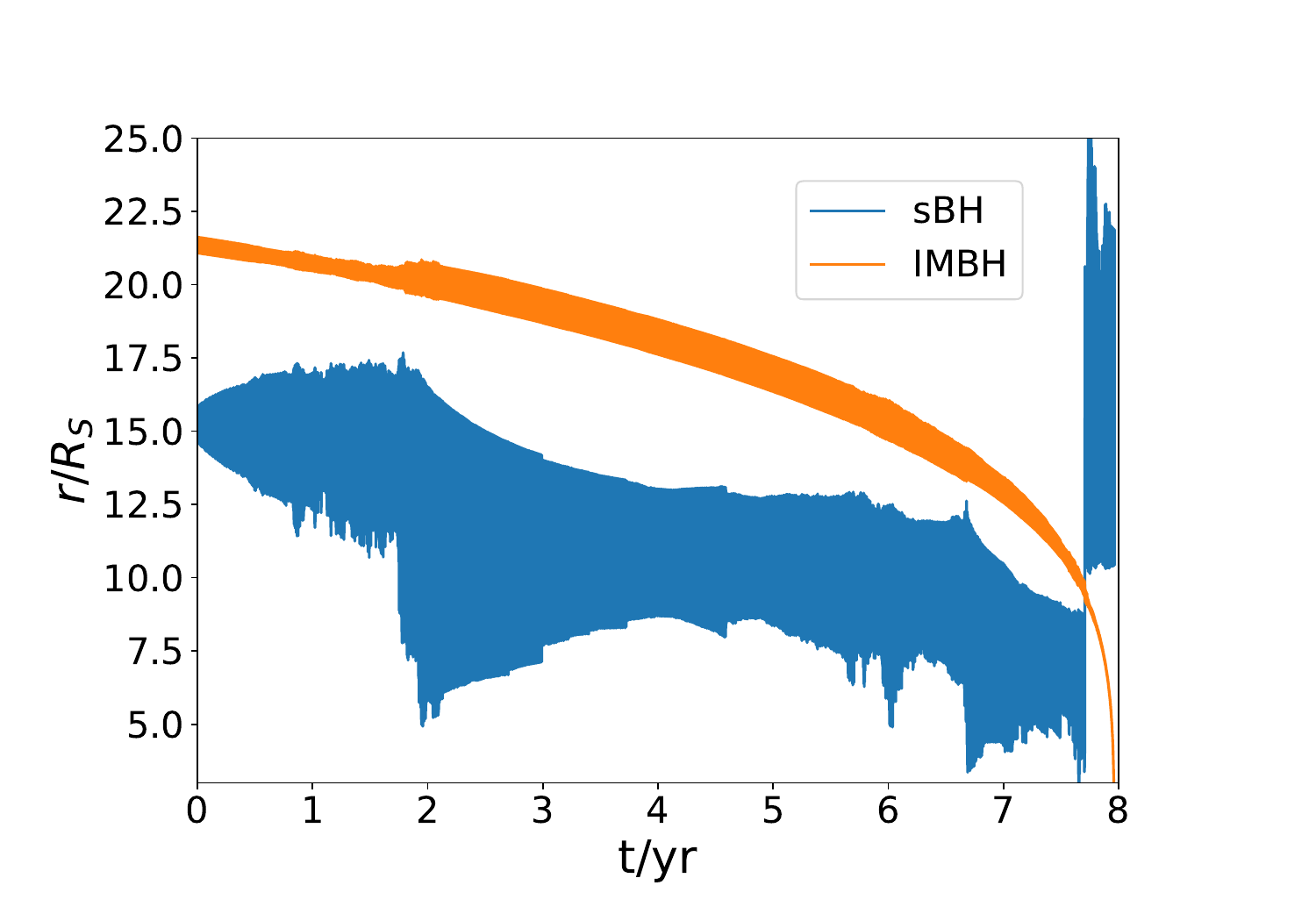}
\caption{Migration of the sBH and the IMBH in the GW dominated scheme, for the case with initial sBH phase angle $0.0 \pi$. The orange and the blue lines represent the orbital radius of the IMBH and the sBH. The x-axis represents the time in unit of yr. The y-axis shows the orbital radius in unit of $R_{\rm{S}}$. } \label{fig:mig_gw_0} 
\end{figure}

One example of the systen evolution is shown in Figure~\ref{fig:mig_gw_0}. The initial sBH phase angle for this case is $0.6\pi$. We can see that initially the sBH is not migrating with the IMBH, since the IMBH is still far away, while the sBH orbital eccentricity is growing due to the perturbation. Around $t= 2$ yr, since the sBH orbit enters one of the mean motion resonances \citep{1999ssd..book.....M} with the IMBH (see Section~\ref{sec:result1} for details), its orbital eccentricity $e_{\rm{sBH}}$ suddenly becomes very large, while $a_{\rm{sBH}}$ decreases significantly, due to the strong tidal torque effect when the mean motion resonance occurs. After that, $e_{\rm{sBH}}$ starts to decrease slowly, due to the orbital eccentricity damping by both the gas and the GW radiation. Around $t= 6.5$ yr, when the orbit of the IMBH is close to the sBH orbit again, similar phenomena take place. So we can see that although the sBH orbital evolution is more irregular in the GW dominated regime, the tidal torque from the IMBH, together with the orbital eccentricity damping effect, can drive the evolution of $a_{\rm{sBH}}$ roughly synchronously with $a_{\rm{IMBH}}$. However, around $t= 7.7$ yr, the IMBH decay by GW radiation is so fast that the tidal torque is not strong enough anymore to synchronize the sBH evolution. Eventually, the IMBH catches up with the sBH inducing strong perturbations in its orbit. 
The orbit of the sBH becomes so chaotic that it intersects with the orbit of the IMBH, swapping the hierarchy of the three body system. The IMBH then eventually merge with the central SMBH, while the sBH is left behind, moving onto an highly eccentric orbit around the MBH.


\begin{figure*}[t]
\centering
\subfigure[\label{fig:mig_gw_k}]{
\includegraphics[width=0.47\textwidth]{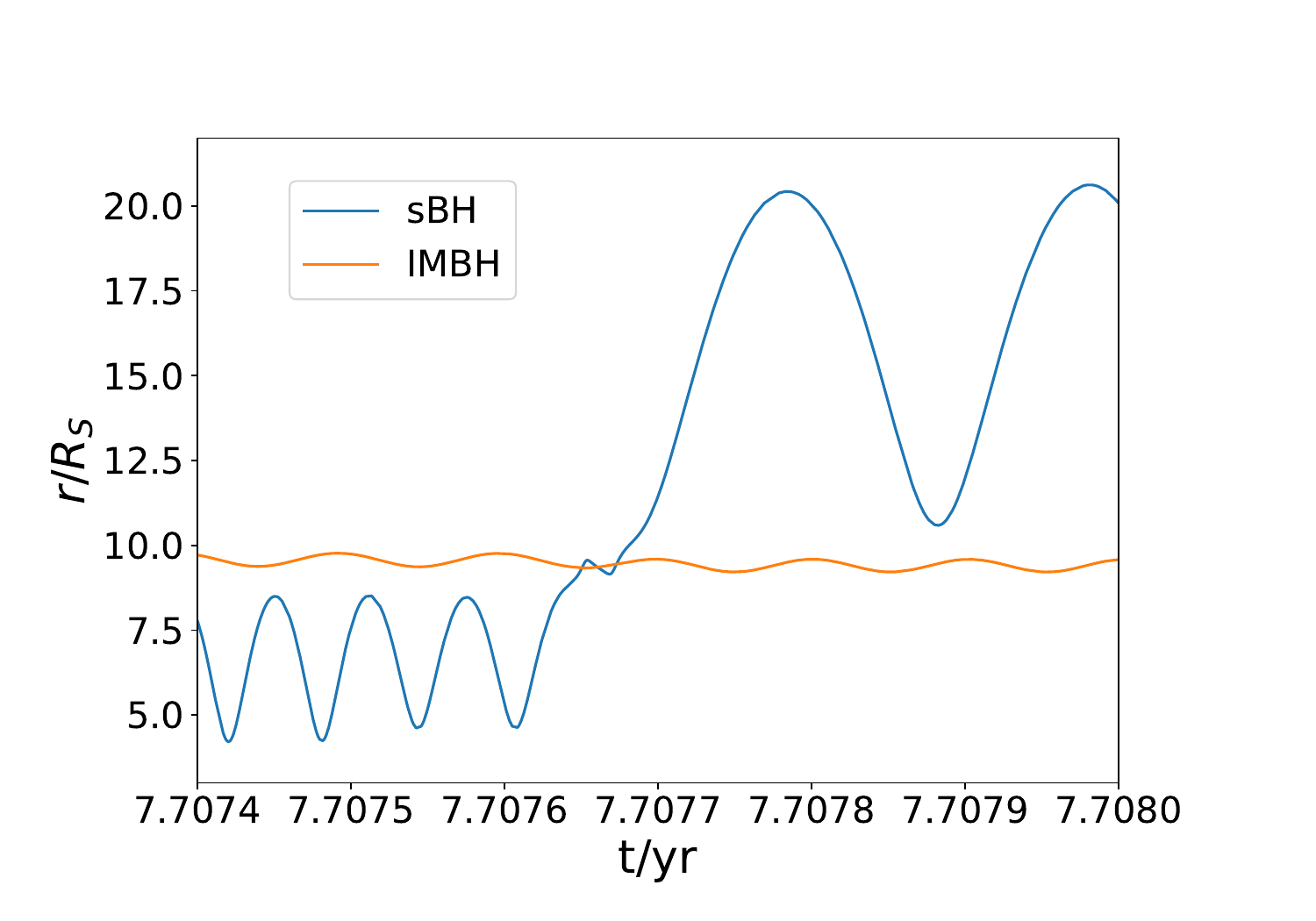}
}
\quad
\subfigure[\label{fig:dis_gw_k}]{
\includegraphics[width=0.47\textwidth]{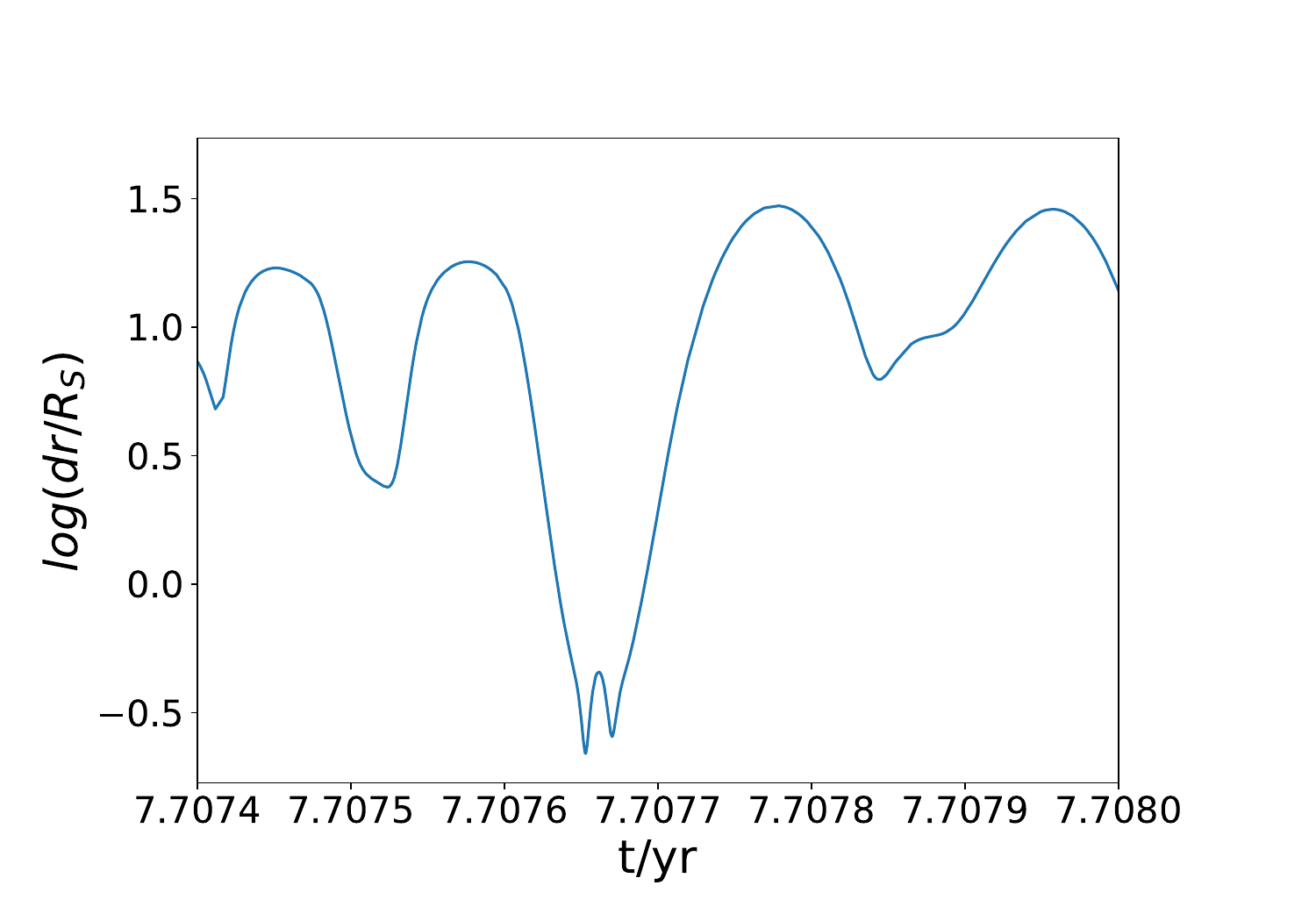} 
}

\caption{(a): Orbital radius of the sBH and the IMBH in the final stage, for the case with initial sBH phase angle $0.0 \pi$. The orange and the blue lines represent the orbital radii of the IMBH and the sBH respectively. The x-axis represents the time in unit of yr. The y-axis shows the orbital radius in unit of $R_{\rm{S}}$.(b): Distance between the sBH and the IMBH in the binary formation stage, for the case with initial sBH phase angle $0.0 \pi$. The x-axis represents the time in unit of yr. The y-axis shows the distance between the sBH and the IMBH in unit of $R_{\rm{S}}$, on logarithmic scale. }
\label{fig:gw_k}
\end{figure*}
 
To better understand the behavior of the sBH when it is 'kicked away' by the IMBH, in Figure~\ref{fig:mig_gw_k} we zoom-in the orbital radius evolution around $t= 7.7$ yr. In Figure~\ref{fig:dis_gw_k}, we instead show the relative distance between the sBH and the IMBH $dr$, verifying the large impact of IMBH gravity on the motion of the sBH. From Figure~\ref{fig:mig_gw_k}, we can see that around one orbital apocenter, the sBH suddenly 'departs' from its original orbit, toward the orbit of the IMBH. The reason is that at this point the distance between the sBH and the IMBH becomes less than the Hills radius of the IMBH, which is $q^{1/3}_{\rm{IMBH}} \, a_{\rm{IMBH}} = 0.1 \, a_{\rm{IMBH}} \sim 1$ \citep{Hills1988}, so the sBH feels the gravitational pull of the IMBH. This close encounter event corresponds to the sudden drop in the distance down to $dr/R_{S} < 1$ in the Figure~\ref{fig:dis_gw_k}. 

Within the Hills radius of the IMBH, the sBH form a binary with the IMBH, corresponding to the stage around $t=7.70763-7.70768$ yr in Figure~\ref{fig:mig_gw_k} and Figure~\ref{fig:dis_gw_k}. Due to the tidal force of SMBH, the binary orbit is irregular, with the closest distance around $dr \lesssim 0.2 $. This binary quickly breaks down, and the sBH is kicked onto an highly eccentric orbit beyond the IMBH. In fact, this kind of transient stage has been broadly studied in planetary dynamics in the so called 'Jacobi capture' problem, in which two planets encounter with each other, with typical impact parameter $\sim$Hills radius, and then form a binary with orbit inside the Hills sphere \citep[e.g.][]{2023MNRAS.518.5653B, 2024ApJ...972..193D}. The binary orbit in the Hills sphere can be chaotic due to the nature of such a three body system, without stable orbital apocenter and pericenter distance. Due to the chaotic nature and the instability of such a system, if there is no dissipation on the binary orbital motion, the system eventually breaks down kicking away one of the planets. The exact lifetime of the binary is also subject to randomness. In our case, the orbital energy of the binary is partially dissipated by the GWs emitted by the sBH-IMBH pair. But the distance of the sBH-IMBH pair is still large enough that the effect on the final outcome of the encounter is negligible, and the sBH escapes from the Hills sphere due to the instability. During the chaotic interaction, the kinetic energy of the IMBH relative to the SMBH is partially transported to the sBH, so after the transient stage the sBH is kicked to an orbit with a much larger orbital radius. In this case, the sBH has experienced a failed Jacobi capture process.

Although the sBH is eventually kicked away, it has already formed a 'temporary' EMRI with the central SMBH which is inside the mHz GW band \citep[e.g. LISA, ][]{{pau18}}, since the orbital radius of the sBH has already migrated to as low as $\sim 5 \, R_{\rm{S}}$. in fact, from Figure~\ref{fig:mig_gw_0}, we can see that the sBH stays in $5$
--$10 \, R_{\rm{S}}$ for about one year. After the sBH is kicked away, the IMBH forms an IMRI, and finally merges with the central SMBH. So in this a case, although the sBH is finally kicked out, we can still possibly detect a 'tempoarary, failed EMRI' and an IMRI in sequence.

\subsection{Outcome 2: Binary formation}

\begin{figure*}[t]
\centering
\subfigure[\label{fig:mig_gw_b}]{
\includegraphics[width=0.47\textwidth]{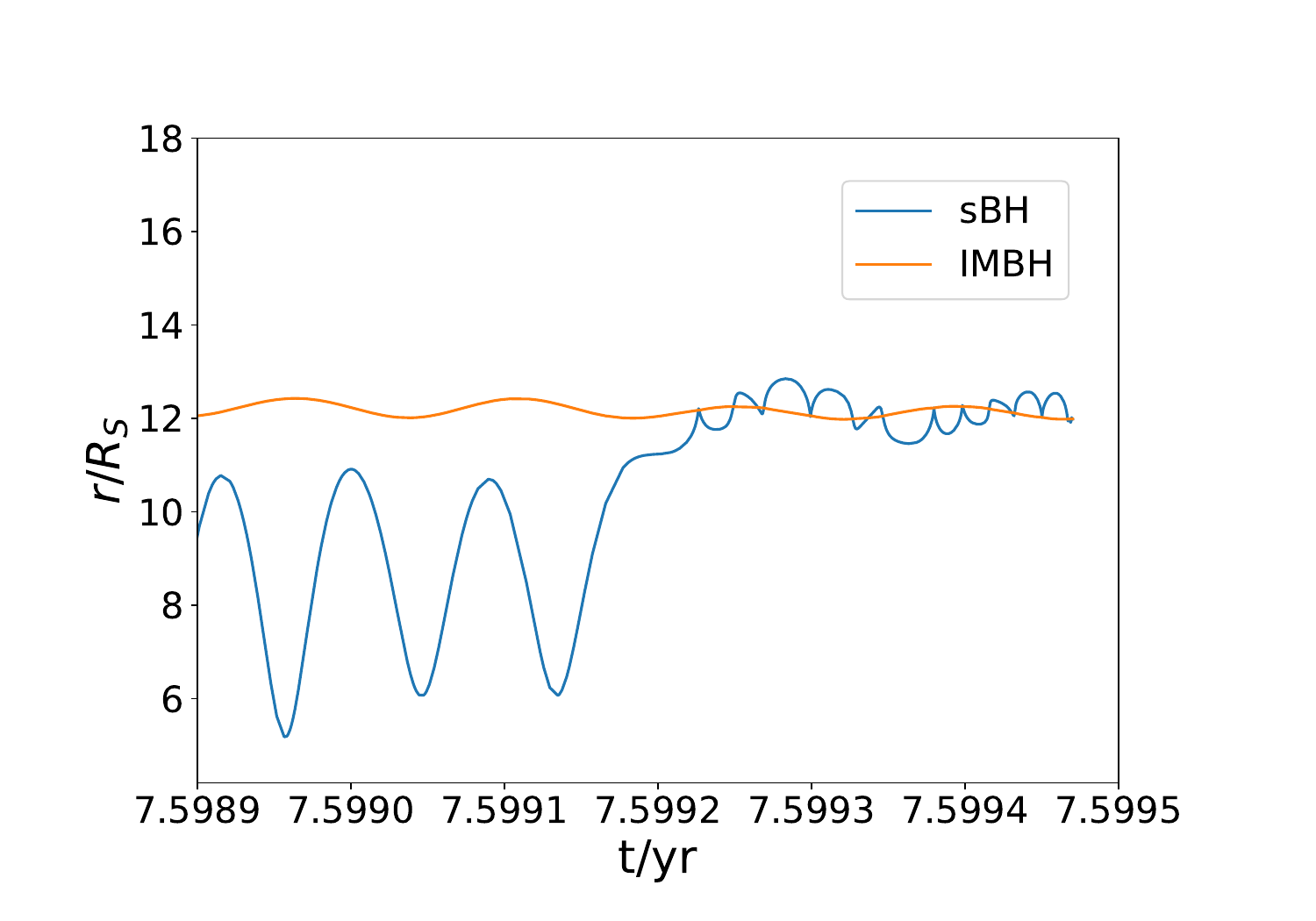}
}
\quad
\subfigure[\label{fig:dis_gw_b}]{
\includegraphics[width=0.47\textwidth]{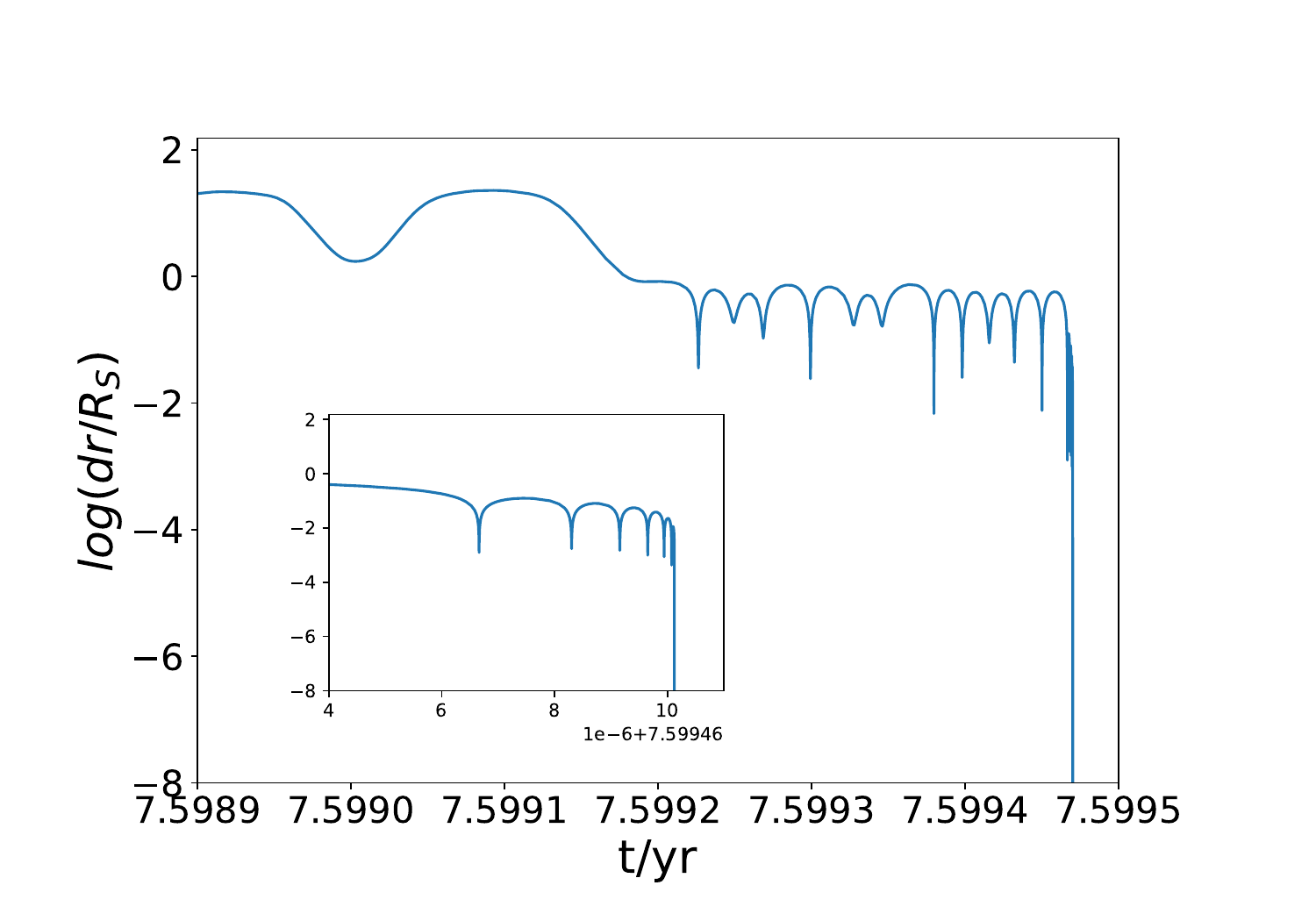} 
}

\caption{(a): Orbital radius of the sBH and the IMBH in the final stage, for the case with initial sBH phase angle $0.6 \pi$. The orange and the blue lines represent the orbital radii of the IMBH and the sBH respectively. The x-axis represents the time in unit of yr. The y-axis shows the orbital radius in unit of $R_{\rm{S}}$.(b): Distance between the sBH and the IMBH in the binary formation stage, for the case with initial sBH phase angle $0.6 \pi$. The x-axis represents the time in unit of yr. The y-axis shows the distance between the sBH and the IMBH in unit of $R_{\rm{S}}$, on logarithmic scale. The inset shows the zoom-in of this figure between $t=7.599464 - 7.599471$ yr. }
\end{figure*}

We examine here a case in which the system forms a stable sBH-IMBH binary induced by the Jacobi capture, followed by the coalescence of the IMBH-sBH binary, thus forming a 'light IMRI'. The initial sBH phase angle for this case is $0.0\pi$.

Similarly to Figure~\ref{fig:gw_k}, we show the evolution of the orbital radii of the sBH and the IMBH, before and during the chaotic three body dynamical interaction, in Figure~\ref{fig:mig_gw_b}. We also show the relative distance between the sBH and the IMBH in Figure~\ref{fig:dis_gw_b}. Similarly to the previous case, once the distance between the sBH to the IMBH is smaller than the Hills radius (around $t=7.5992$ yr), the sBH is 'diverted' toward the IMBH and forms a binary with it. In this stage, the motion of the sBH is very chaotic, without a stable apocenter and pericenter distance, similar to what we saw in the last section. 

However, in this case, the minimum distance between the sBH and the IMBH is as low as $dr \sim 0.01$ ( versus $\sim 0.2$ in the previous case). 
This corresponds to $\sim 10$ times Schwarzschild radius of the IMBH, and the associated GW radiation is strong enough to dissipate the orbital energy within just few encounters. So after such encounters, the orbit of sBH-IMBH binary becomes stable and we have a successful Jacobi capture process producing a genuine sBH-IMBH binary, i.e. a 'light IMRI'.

The distance between the sBH and the IMBH after the stable binary formation is shown in the Figure~\ref{fig:dis_gw_b} inset. We can see that compared to the motion in the transient stage, the orbit in this stage is more regular, showing stable apocenter and pericenter passages. 
Due to the very strong GW radiation, the sBH quickly inspiral into the IMBH in just few orbits. Due to restriction of the simulation setup, the code cannot continue to calculate the motion of the IMBH after the merger of the sBH-IMBH binary. But since the IMBH is already very close to the SMBH ($\sim 10 \, R_{\rm{S}}$), it will then inspiral into the SMBH within about one year. So in this case, two successive IMRI events, i.e. the sBH-IMBH binary and IMBH-SMBH binary, with different chirp masses, could be detected by the GW detector within one year. 


\subsection{Statistical result}

\begin{figure*}[t]
\centering
\includegraphics[width=1.0\textwidth]{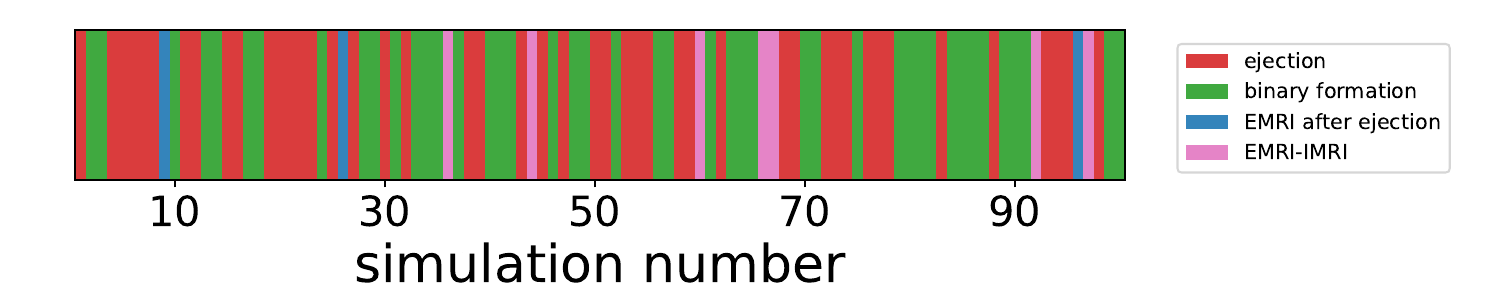}
\caption{ Distribution of different kinds of outcomes for different simulations. The red, green, blue and pink strips represent the outcome of ejection, binary formation, sBH still within $\sim 10 \, R_{\rm{S}}$ after ejection and complete EMRI before IMRI.}
\label{fig:mig_gw_stat_2} 
\end{figure*}

In the above two cases, all the initial conditions are the same, except the initial phase angle of the sBH. Still, the long term evolution, especially the dynamical behavior of the sBH during and after it is captured in the Hills sphere of the IMBH, is very different. This highlights that the dynamical evolution and the final fate of the sBH is very sensitive to the initial condition due to the chaotic nature of the three body problem. Specifically, different initial phase angle of the sBH will result in different impact parameter when the sBH first enters into the Hills sphere, which can further induce a very different set of close approaches and lifetime in the transient stage. 

\begin{figure*}[t]
\flushleft
\includegraphics[width=0.95\textwidth]{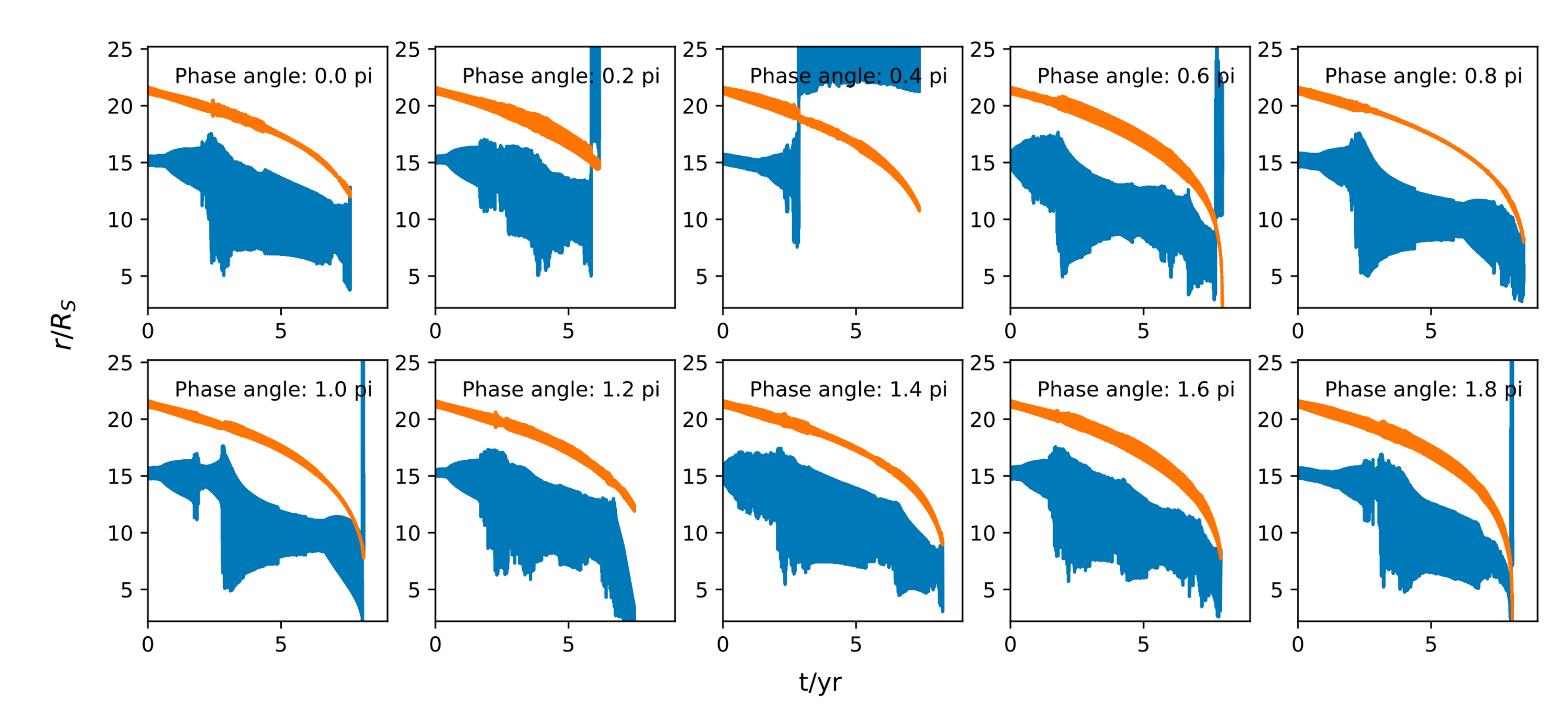}
\caption{ Ten examples of the sBH and IMBH orbital evolution in the GW dominated scheme, with sBH initial phase angle as 0.0, 0.2, 0.4, 0.6, 0.8, 1.0, 1.2, 1.4, 1.6 and 1.8 $\pi$. The orange and blue lines represent the orbital radius of IMBH and sBH, respectively. The x-axis represents the time in unit of yr. The y-axis shows the orbital radius in unit of $R_{\rm{S}}$. The text in every sub-figure represents the initial phase angles of the sBH. }
\label{fig:mig_gw_stat_1}
\end{figure*}

To explore the probability of the different outcomes (swap and sBH ejection versus sBH-IMBH binary formation), we perform a set of simulations with different initial phase angles. More specifically, the initial phase angle of the IMBH is kept as 0, while the initial phase angle of the sBH is chosen in $0-2\pi$ with interval $0.02\pi$. The other initial conditions, such as the semi-major axis and the eccentricity, are kept the same. We ran a total of 100 simulations, with the final outcomes summarized in the Figure~\ref{fig:mig_gw_stat_2}. Different colors represent different kind of outcomes as labeled. To highlight the different kind of evolutions, in Figure~\ref{fig:mig_gw_stat_1}, we show the radial evolution for ten of them, showing the migration of the IMBH and the sBH, as well as the final fate of the sBH. The orange line and the blue line represent the orbital radius of the IMBH and the sBH, respectively. The numbers of different outcomes in Figure~\ref{fig:mig_gw_stat_1} are roughly consistent with the statistic for all 100 simulations.

Considering the whole set of 100 simulations, about 44 percent of the simulations end with binary formation and merger of the light IMRI, shown by the green strips in Figure~\ref{fig:mig_gw_stat_2}. They are represented by the case with initial phase angle of $0.0\pi$, $0.8\pi$, $1.4\pi$ and $1.6\pi$, in Figure~\ref{fig:mig_gw_stat_1}. In all of these cases, the light IMRI merges within a few orbits after the binary formation. The light IMRI formation and merger happen when the IMBH migrates to $\lesssim 15 \, R_{\rm{S}}$, so after the merger the IMBH will fall into the SMBH within $\sim 3$ yr \citep[merger timescale,][]{maggiore2008gravitational}. In this kind of situations, the GW detectors can detect two successive IMRI events with different chirp mass ($\sim 10^2$ and $10^4 \, M_\odot$, respectively). We also point out that the GW waveform of the light IMRI could be very different from an isolated IMRI, since the merger event happens very close to the central SMBH \citep[e.g.][]{2019PhRvD.100f3012T,2020PhRvD.101h3031D,2023MNRAS.521.2919Z,2024ApJ...968..122Z}.

Conversely, in 46 percent of the all cases, the sBH and the IMBH has not merged finally, shown by the red strips in Figure~\ref{fig:mig_gw_stat_2}. The sBH is ejected after a transient binary stage. These events are represented by the cases with initial phase angle as $0.2\pi$, $0.4\pi$, $0.6\pi$, $1.0\pi$ and $1.8\pi$ in Figure~\ref{fig:mig_gw_stat_1}. In most of the ejection cases, one temporary EMRI event will happen before the IMRI, since the orbit of the sBH has already enters the mHz GW band ($\sim 10 \, R_{\rm{S}}$) before being kicked out. Then the IMBH will inspiral into the SMBH within a few years for these cases. Thus, a mHz GW detector could also detect a 'temporary failed EMRI' and an IMRI in sequence. In only $\sim$4 percent of all cases, the sBH is kicked out too early so that only a single IMRI will happen. An example is shown by the case of $0.4 \pi$ in Figure~\ref{fig:mig_gw_stat_1}, in which the sBH is kicked out around $\sim 20 \, R_{\rm{S}}$.


Finally, there are some exotic cases not included in the poll discussed above. For example, in $\sim 7$ percent of all simulations, the sBH just falls into the SMBH before being kicked or captured, as shown by the pink strips in Figure~\ref{fig:mig_gw_stat_2}. One of these cases is shown in Figure~\ref{fig:mig_gw_stat_1}, corresponding to a phase angle of $1.2 \pi$. In this kind of cases, there will be an complete EMRI event followed by an IMRI event. In another $\sim 3$ percent cases, although the sBH is kicked out, the orbital radius of the sBH is still not very large after ejection (e.g. $\sim 10 \, R_{\rm{S}}$). These cases are shown by the blue strips in Figure~\ref{fig:mig_gw_stat_2}. So in these cases, there will be an IMRI event followed by an EMRI event.

We have seen that in most cases two GW events can be detected successively, either two IMRIs, EMRI-IMRI or IMRI-EMRI. In less than half of these cases, the sBH-IMBH binary can form. So the probability of the sBH-IMBH binary formation is of the order of $0.1$, which is roughly consistent with the previous works on the binary formation in the AGN disk. As mentioned before, the binary formation process shown in this work can be described by the Jacobi capture scenario, in which a sBH-IMBH binary with irregular motion can form when the sBH enter into the Hills sphere. Whether the binary could become stable and merge, depends on whether the energy dissipation due to the GW radiation is strong enough. Since the strength of the GW radiation is severely dependent on the distance between the sBH and the IMBH \citep{maggiore2008gravitational}, so the final result will be roughly determined by the closest distance between the sBH and the IMBH during the transient stage. To reduce the orbital energy in several binary orbital timescale, the closest distance between the sBH and the IMBH should be as low as $\sim 0.01$ (also $\sim 0.01$ times the Hills sphere) \citep[e.g.][]{Li2022_binaryform}. According to the previous works, the probability that the closest distance is smaller than 0.01, can be $\lesssim 0.1$ \citep[e.g.][]{2023MNRAS.518.5653B, 2024ApJ...972..193D}, which should also be the rough binary formation probability. In our case, the binary formation probability is slightly higher, which may result from the minor contribution on the energy dissipation by the gaseous torque. 

\section{Conclusion}

\label{sec:conclusion}

In this work, we extended the work in Paper I by investigating the interaction between a gap-opening IMBH and a surrounding sBH in the AGN disk. Using 3D hydrodynamical simulations, we found that the migration of sBHs around the inner edge of the gap opened by the IMBH can be significantly accelerated by both gaseous and tidal torques. We further found that, although the inner disk is partially accreted, the gaseous torque together with the tidal torque of IMBH can induced synchronized migration until approximately ten Schwarzschild radii from the central SMBH. 

We further used a relativistic (PN motion) three-body simulation setup to study the final fate of the sBH in the GW-dominated regime and found that in most cases the sBH can be either captured or ejected by the IMBH. In both cases, two successive GW events can happen. For the ejection case, a temporary EMRI will enter the LISA band, followed by an IMRI event. For the capture case, two IMRIs will happen successively: the first forms from the sBH-IMBH binary, the second one corresponds to the IMBH inspiralling into the SMBH, with a time interval $\lesssim 3$ year.

Although we only run simulations with the IMBH mass $m_{\rm{IMBH}} = 1000 \, M_{\odot}$ due to the resolution limitation, our conclusions should also apply to less massive IMBHs, provided $m_{\rm{IMBH}}$ is sufficient to open even a shallow gap in the disk. In the initial stage, where GW radiation is not significant, both the gaseous and tidal torques can still drive synchronized migration, even for cases with a smaller gap-opening IMBH. For example, the Type-I torque will be effectively strengthened as long as the disk profile is perturbed by the IMBH. Also, the interfering density wave and tidal torque will always be effective when the orbital timescale ratio between the sBH and the IMBH is approximately an integer. In the second stage, where GW radiation dominates, a smaller IMBH will result in slower migration, generally enhancing the stability of synchronized migration. The ejection or capture process should still occur, but in a region closer to the SMBH, which makes it even easier to produce two successive EMRI/IMRI events. 

Estimating the event rate for such successive EMRI/IMRI events is challenging due to many theoretical uncertainties, including the number of sBHs within an AGN accretion disk and the formation channels of IMBHs. If the IMBHs are produced by successive mergers of the sBHs inside migration traps, we can estimate the event rate in the following way. Recent theoretical models, which do not account for migration traps, predict an EMRI event rate in AGNs of $10$ to $10^4$ per year \citep{pan21b}. If a migration trap is present in the disk, the sBHs will no longer form EMRIs but accumulate in the trap. Since on average $10-100$ sBHs are needed to form a gap-opening IMBH which can leave the migration trap, the formation rate of the successive EMRI/IMRI events would be about $1$--$10^2$ per year in this scenario.

\section*{Acknowledgements}

This work is supported by the National Key Re- search and Development Program of China (Grant No. 2021YFC2203002) and the National Natural Science Foundation of China (Grant No. 12473037). 
AF acknowledges support provided by the "GW-learn" grant agreement CRSII5 213497 and the Tomalla Foundation.


\bibliographystyle{aasjournal}
\bibliography{bibbase,mybib,References}{}
\end{document}